\documentclass[sigconf,natbib=true]{acmart}

\usepackage{float} 
\usepackage{bm}
\usepackage{graphicx}
\usepackage{booktabs}
\usepackage{multirow}
\usepackage{enumitem}
\setlist[itemize]{leftmargin=*}
\setlist[enumerate]{leftmargin=*}
\definecolor{lightgray}{RGB}{215,215,215}
\definecolor{lightblue}{RGB}{107,174,214}
\definecolor{bluu}{HTML}{ECF4FF}
\definecolor{blu}{RGB}{158,202,225}
\definecolor{myorange}{RGB}{2, 142, 2}
\usepackage{colortbl}  
\usepackage{color}
\usepackage{xcolor}
\usepackage[normalem]{ulem}
\useunder{\uline}{\ul}{}
\usepackage{subfigure}
\usepackage{wrapfig}
\usepackage{amsmath}

\newcommand{\sss}[1]{\subsubsection{\textbf{#1}}}

\newcommand{\ie}{\emph{i.e., }}
\newcommand{\eg}{\emph{e.g., }}

\newcommand{\etc}{\emph{etc.}}

\copyrightyear{2025}
\acmYear{2025}
\setcopyright{cc}
\setcctype{by}
\acmConference[KDD '25]{Proceedings of the 31st ACM SIGKDD Conference on Knowledge Discovery and Data Mining V.2}{August 3--7, 2025}{Toronto, ON, Canada}
\acmBooktitle{Proceedings of the 31st ACM SIGKDD Conference on Knowledge Discovery and Data Mining V.2 (KDD '25), August 3--7, 2025, Toronto, ON, Canada}
\acmDOI{10.1145/3711896.3736919}
\acmISBN{979-8-4007-1454-2/2025/08}

\begin{document}

\title{EARN: Efficient Inference Acceleration for LLM-based Generative Recommendation by Register Tokens}

\author{Chaoqun Yang}
\email{chaoqun@yang.email.cn}
\affiliation{
\institution{Tsinghua University}
\city{Beijing}
\country{China}
}

\author{Xinyu Lin}
\authornote{Corresponding authors. This research is supported by the National Research Foundation, Singapore under its National Large Language Models Funding Initiative (AISG Award No: AISG-NMLP-2024-002).
Any opinions, findings and conclusions or recommendations expressed in this material are those of the author(s) and do not reflect the views of National Research Foundation, Singapore.}
\email{xylin1028@gmail.com}
\affiliation{
\institution{National University of Singapore}
\city{Singapore}
\country{Singapore}
}

\author{Wenjie Wang}
\email{wenjiewang96@gmail.com}
\authornotemark[1]
\affiliation{
\institution{University of Science and Technology of China}
\city{Hefei}
\country{China}
}

\author{Yongqi Li}
\email{liyongqi0@gmail.com}
\affiliation{
\institution{The Hong Kong Polytechnic University}
\city{Hong Kong}
\country{China}
}

\author{Teng Sun}
\email{stbestforever@gmail.com}
\affiliation{
\institution{Shandong University}
\city{Qingdao}
\country{China}
}

\author{Xianjing Han}
\email{hanxianjing2018@gmail.com}
\affiliation{
\institution{National University of Singapore}
\city{Singapore}
\country{Singapore}
}

\author{Tat-Seng Chua}
\email{dcscts@nus.edu.sg}
\affiliation{
\institution{National University of Singapore}
\city{Singapore}
\country{Singapore}
}

\renewcommand{\shortauthors}{Chaoqun Yang et al.}

\begin{abstract}
Large Language Model-based generative recommendation (LLMRec) has achieved notable success, but it suffers from high inference latency due to massive computational overhead and memory pressure of KV Cache.
Existing KV Cache reduction methods face critical limitations: cache compression offers marginal acceleration given recommendation tasks' short decoding steps, while prompt compression risks discarding vital interaction history. 
Through systematic analysis of attention patterns in LLMRec, we uncover two pivotal insights: 1) \textit{layer-wise attention sparsity inversion} where early layers retain dense informative patterns while later layers exhibit high redundancy, and 2) \textit{dual attention sinks phenomenon} where attention scores concentrate on both head and tail tokens of input sequences.
Motivated by these insights, we propose \textbf{EARN}, an efficient inference framework that 
leverages the early layers to compress information into register tokens 
placed at the input sequence boundaries, then focuses solely on these tokens in the subsequent layers. 
Extensive experiments on three datasets, two LLMRec methods and two LLM architectures demonstrate EARN's superiority, achieving up to 3.79x speedup and 80.8\% KV Cache reduction with better accuracy than the general finetuning approach. 
Our work bridges the efficiency-effectiveness gap in LLMRec, offering practical deployment advantages for industrial scenarios.

\end{abstract} 




\maketitle

\newcommand\kddavailabilityurl{https://doi.org/10.5281/zenodo.15553291}
\ifdefempty{\kddavailabilityurl}{}{
\begingroup\small\noindent\raggedright\textbf{KDD Availability Link:}\\
The source code of this paper has been made publicly available at \url{\kddavailabilityurl}.
\endgroup
}

\section{Introduction}

Recently, Large Language Model (LLM)-based generative recommendation (LLMRec) has shown great potential due to its superior reasoning capabilities \cite{lin2025can,wang2023generative,wu2024survey,kim2024large,lin2024bridging,lin2024data,li2024survey}. 
However, it suffers from high inference latency due to its massive model architecture and auto-regressive decoding paradigm, which severely limits its application in practical recommendation scenarios \cite{zhou2024survey,li2023large,wu2024efficient_survey}. 
For instance, platforms like YouTube, which serve hundreds of millions of users daily with billions of interactions, require millisecond-level response times\footnote{https://affmaven.com/youtube-statistics/}.
This highlights the stark efficiency gap between current LLMRec implementations and real-world application requirements.
Therefore, enhancing the inference efficiency of LLMRec is a crucial issue for industrial applications.

\begin{figure}[t]
    \vspace{-0.2cm}
    \setlength{\abovecaptionskip}{0cm}
    \setlength{\belowcaptionskip}{-0.3cm}
    \centering
    \includegraphics[width=1\linewidth]{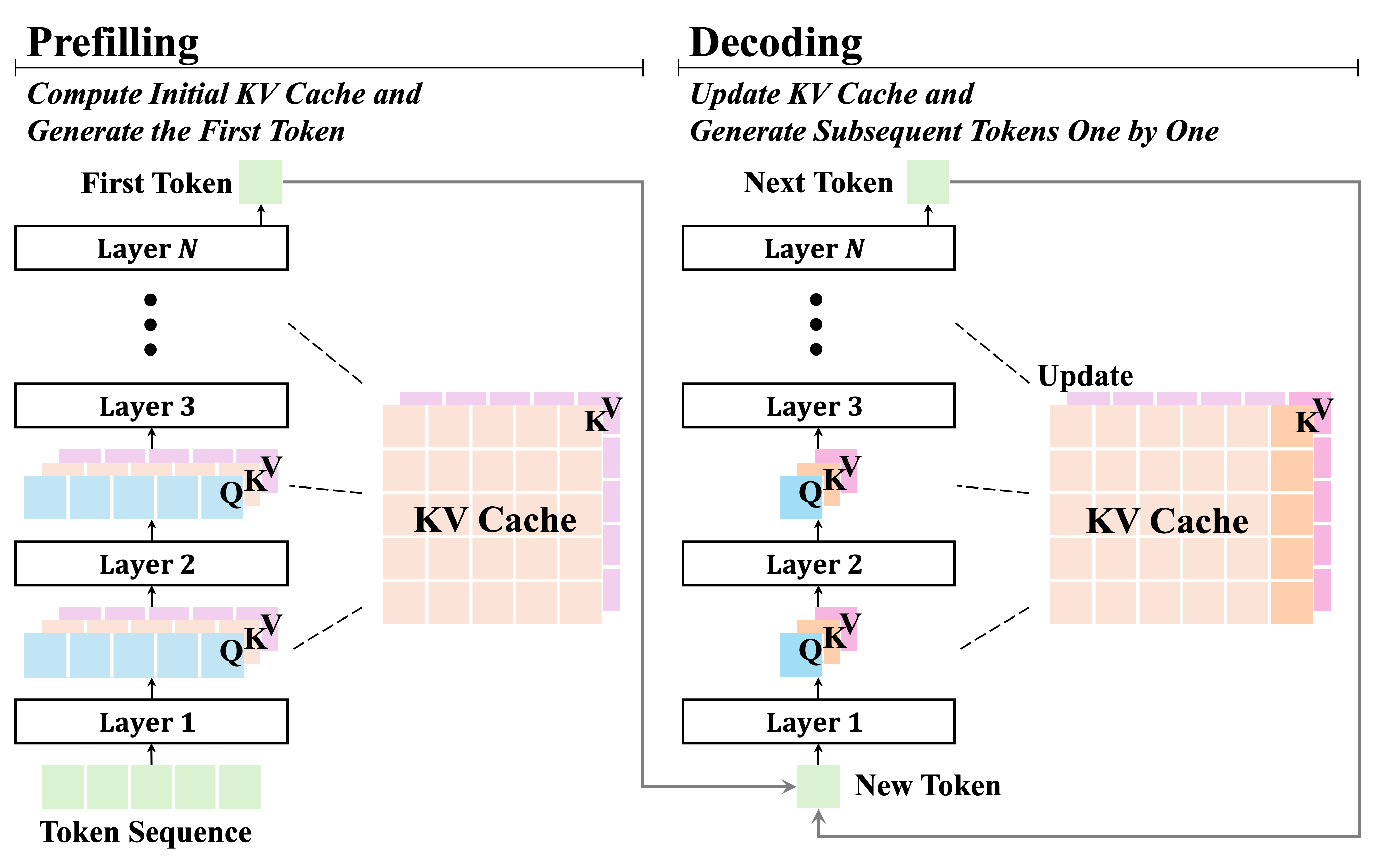}
        \caption{The inference process of LLMRec. In the \textit{prefilling} stage, LLM computes the initial KV Cache and generates the first token of the output; in the \textit{decoding} stage, LLM updates the KV Cache and generates subsequent tokens one by one.}
    \label{fig:LLM-inference}
\end{figure}

For efficient inference of LLMRec, caching computed Key-Value state pairs (\ie KV Cache) is a popular choice. Technically, the inference pipeline with KV Cache typically consists of two distinct stages. 
As shown in Figure~\ref{fig:LLM-inference}, given the input token sequence (user profile, historical interactions, \etc), the first is the \textit{prefilling} stage, where LLM generates the first token of the output and caches computed Key-Value state pairs of all layers for every token. 
The second is the \textit{decoding} stage, where LLM generates subsequent tokens auto-regressively using pre-computed KV Cache, and updates KV Cache in each decoding step. 
As the input sequence length increases, the computational overhead of KV Cache grows greatly, significantly prolonging the inference latency.
And when the KV Cache size is large, each decoding step requires accessing a substantial amount of memory, which can severely slow down the inference.
In light of the above, the key to accelerating the inference of LLMRec lies in reducing the size of KV Cache.

To address this challenge, some research has been proposed. One of the representative works is cache compression~\cite{zhou2024survey,shi2024keep}. It selectively removes less important KV pairs to reduce KV Cache size during \textit{decoding}. 
However, recommendation tasks typically require few decoding steps to generate short output token sequences, such as item identifiers (around 1 to 5 tokens), limiting their acceleration potential in \textit{decoding}. Another method is prompt compression~\cite{zhou2024survey,li2024prompt}, which reduces initial KV Cache size during \textit{prefilling} by condensating the input sequences, for instance, deleting unimportant tokens~\cite{li2023compressing} or compressing the token sequence into a few tokens~\cite{li2024500xcompressor}.
This technique not only decreases the computational load in \textit{prefilling} but also alleviates the memory pressure in \textit{decoding}. 
Although this method has achieved success in NLP~\cite{zhou2024survey,li2024prompt}, it struggles to balance efficiency and accuracy in the context of LLMRec. In LLMRec, it is difficult to distinguish between important and unimportant items, which makes prompt compression method prone to discard crucial user interaction information, resulting in severe accuracy losses, or fail to achieve satisfactory acceleration in an effort to retain information. 
Hence, there is an urgent need for a method that enhances inference efficiency while preserving recommendation effectiveness.

\begin{figure}[t]
    \vspace{-0.2cm}
    \setlength{\abovecaptionskip}{0cm}
    \setlength{\belowcaptionskip}{-0.3cm}
    \centering
    \includegraphics[width=1\linewidth]{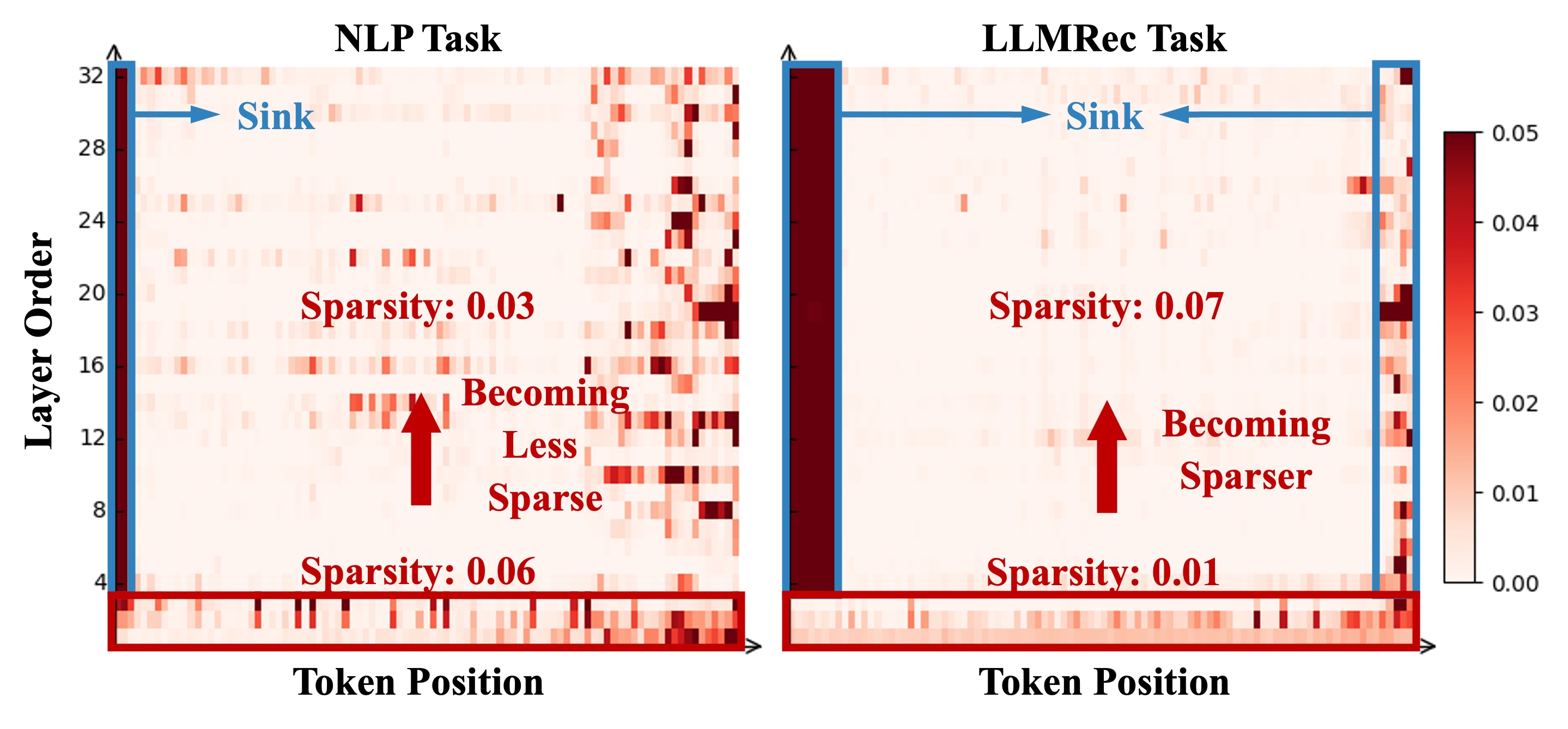}
    \caption{The attention distributions of different layers on the NLP task (reading comprehension QA) and the LLMRec task (LC-Rec on Beauty dataset) on Head 0 of Llama model. In the LLMRec task, the first three layers are relatively dense, while subsequent layers are sparse, with sinking occurring on the head and tail tokens.}
    \label{fig:attention}
\end{figure}

To this end, we identify which information can be safely compressed in LLMRec, by inspecting the attention score distributions~\cite{xiao2023efficient} in LLMRec.
Along two critical dimensions: layer order and token position, we observe two key characteristics of LLMRec across three diverse datasets, two mainstream LLMRec methods, and two distinct LLM architectures. 
For example, as shown in Figure~\ref{fig:attention}, on the Llama model and Beauty dataset, we found the following (more similar evidence can be found in Appendix~\ref{app:attention}):
\begin{itemize}[leftmargin=*]
\item \textbf{Layer-wise attention sparsity inversion.} 
Along the layer order dimension, we observe distinct sparsity between the NLP task and the LLMRec task. In the NLP task, the attention score distribution is sparse in the initial layers, but becomes relatively less sparse in the later layers (Sparsity: $0.06 \rightarrow 0.03$). Conversely, in the LLMRec task, the initial layers exhibit a relatively denser attention distribution, while the subsequent layers are highly sparse (Sparsity: $0.01 \rightarrow 0.07$).
\label{sec:sink}
\item \textbf{Dual attention sinks phenomenon.} 
Along the token position dimension, while both the NLP task and the LLMRec task exhibit attention sinks~\cite{xiao2023efficient}, which means attention scores are highly concentrated on specific token positions, their distributions differ significantly. In the NLP task, sinks primarily emerge at the initial positions and a broad range of later tokens. In contrast, LLMRec demonstrates a distinctive dual concentration pattern - strong attention sinks at both head positions and a narrow tail region.
\end{itemize}
These findings suggest that in LLMRec: \textbf{1)} In terms of the layer order, the early layers contain richer information, while the middle tokens in the later layers are redundant. In fact, previous studies have shown that the early layers of LLMs are capable of summarizing required information and important tokens~\cite{shi2024discovering}, while the later layers tend to be increasingly redundant, and pruning of redundant parts has minimal impact on performance~\cite{li2024mix}. \textbf{2)} In terms of the token position, the head and tail tokens carry the most important information. As explained in prior studies~\cite{xiao2023efficient,gu2024attention}, head tokens can absorb excess attention. Meanwhile, tail tokens interact with all previous tokens, which suggests that tail tokens of the prompt may possess the potential to summarize the preceding user interaction information.

Inspired by the above observations, we propose \textbf{EARN}, an \textbf{E}fficient inference \textbf{A}cceleration method for LLM-based \textbf{R}ecommendation by register toke\textbf{n}s (Figure~\ref{fig:method}), to enhance inference efficiency while maintaining recommendation effectiveness. 
Specifically, EARN introduces prefix and suffix register tokens - several learnable virtual tokens placed at the beginning and end of the input sequence, and leverages the first \textit{k} layers of LLM to compress the user's historical interaction information into register tokens.
During inference, we use only register tokens for computation after \textit{k} layers, thereby reducing the computational load and memory footprint of the KV Cache.
We instantiate EARN on two mainstream LLMRec methods with two distinct LLM architectures, and conduct extensive experiments on three real-world datasets, validating the superiority of EARN in terms of both efficiency and accuracy. The code and datasets are available at \url{https://github.com/transcend-0/EARN}.

The contributions of our work are manifold:
\begin{itemize}[leftmargin=*]
    \item We identify the layer-wise attention sparsity inversion and dual attention sinks phenomenon in LLMRec, revealing that early layers as well as both head and tail tokens, retain the most critical information for LLMRec. 
    \item We propose an efficient and effective method that leverages register tokens to compress user historical interactions and prune redundant computations in later layers, striking a delicate balance between inference efficiency and recommendation effectiveness.
    \item We validate the effectiveness of EARN through extensive experiments conducted on three real-world datasets, demonstrating the superiority of EARN in achieving both high inference speed and recommendation accuracy.
\end{itemize}

\section{Preliminary}

\subsection{LLM-based Generative Recommendation}
In this work, we primarily discuss the LLM-based generative recommendation. The main idea of LLM-based generative recommendation is using LLMs as the core recommender model, \ie taking the task instruction and a user's historical interactions as the input prompt to generate item identifiers, such as typical item IDs.
Formally, given a user's historical interactions $H=(i_1,\cdots,i_T)$ in the chronological order, where $i_1$ to $i_T$ represent the items the user has interacted with in the past, LLM-based generative recommendation predicts the next item $i_{T+1}$ the user is likely to interact with, \ie
\begin{equation}
f(H)=P(i_{T+1}\ |\ i_1,\cdots,i_T) \ ,
\end{equation}
where $f$ is the LLM, $P$ is the probability distribution of the items.

\subsection{Inference of LLMRec}

LLM-based generative recommendation follows an auto-regressive decoding paradigm to generate item identifiers, which comprises two distinct computational stages: \textit{prefilling} and \textit{decoding}. Each stage exhibits unique inference characteristics and optimization requirements.

\sss{Prefilling}
In this stage, the input prompt (containing the task instruction and the user's historical interactions) is tokenized into the token sequence and fed into the transformer decoder. The most pivotal component is the multi-head attention module that exists in each layer of the model. 
Within each attention head, the input hidden states $\bm{X} \in \mathbb{R}^{n \times d}$ ($n$ denotes the sequence length, and $d$ denotes the hidden dimension) undergo linear transformations:
\begin{equation}\small
\bm{Q} = \bm{XW_Q}, \quad \bm{K} = \bm{XW_K}, \quad \bm{V} = \bm{XW_V} \ .
\end{equation}
Subsequently, the attention mechanism computes the attention output as follows:
\begin{equation}\small
\bm{O}=Softmax\left(\frac{\bm{Q}\bm{K}^T}{\sqrt{d}}\right)\bm{V} \ .
\end{equation}
where $\bm{O} \in \mathbb{R}^{n\times d}$ is the attention output, $\bm{Q}, \bm{K}, \bm{V} \in \mathbb{R}^{n\times d}$ correspond to query states, key states, and value states, respectively.
Since the generation of subsequent tokens utilizes key states $\bm{K}$ and value states $\bm{V}$ of preceding tokens, it is common practice to cache $\bm{K}$ and $\bm{V}$ for subsequent use, \ie KV Cache. This caching strategy is crucial for maintaining the efficiency of the inference process.

During the \textit{prefilling} stage, $\bm{Q}$, $\bm{K}$ and $\bm{V}$ are all matrices, and matrix-matrix multiplication is compute-bound, meaning that the inference speed is primarily determined by the number of floating-point operations (FLOPs). Reducing the computational load can effectively accelerate the \textit{prefilling} stage. For instance, shortening the input token sequence length can help decrease the quadratic computational complexity $O(n^2d)$ of the attention mechanism. Besides, a shorter token sequence also reduces the size of KV Cache, thereby contributing to accelerating the \textit{decoding} stage.

\sss{Decoding}
In this stage, subsequent tokens are generated one by one, with the KV Cache being expanded accordingly. 
At each decoding step $t$, the model first computes $\bm{Q}_t, \bm{K}_t, \bm{V}_t \in \mathbb{R}^{1 \times d}$, then updates the KV Cache:  
   \begin{equation}
   \bm{K}_{1:t} = Concat(\bm{K}_{1:t-1}, \bm{K}_t), \quad \bm{V}_{1:t} = Concat(\bm{V}_{1:t-1}, \bm{V}_t) \ .
   \end{equation}
Next, the attention output is calculated via $\bm{Q}_t\bm{K}_{1:t}^T$. 

Unlike the \textit{prefilling} stage, during the \textit{decoding} stage, $\bm{Q}$ is a vector, while $\bm{K}$ and $\bm{V}$ are matrices. The vector-matrix multiplication is memory-bound, meaning that the inference speed is primarily limited by the efficiency of memory access rather than the computational capacity. The latency is thus predominantly determined by the efficiency of memory access to the growing KV Cache rather than FLOPs.
In other words, reducing the size of KV Cache allows the GPU computing cores to access them more rapidly, which is the key to accelerating inference in the \textit{decoding} stage.

\section{Method}

\begin{figure}[!t]
    \vspace{-0.2cm}
    \setlength{\abovecaptionskip}{0cm}
    \setlength{\belowcaptionskip}{0cm}
    \centering
    \includegraphics[width=\linewidth]{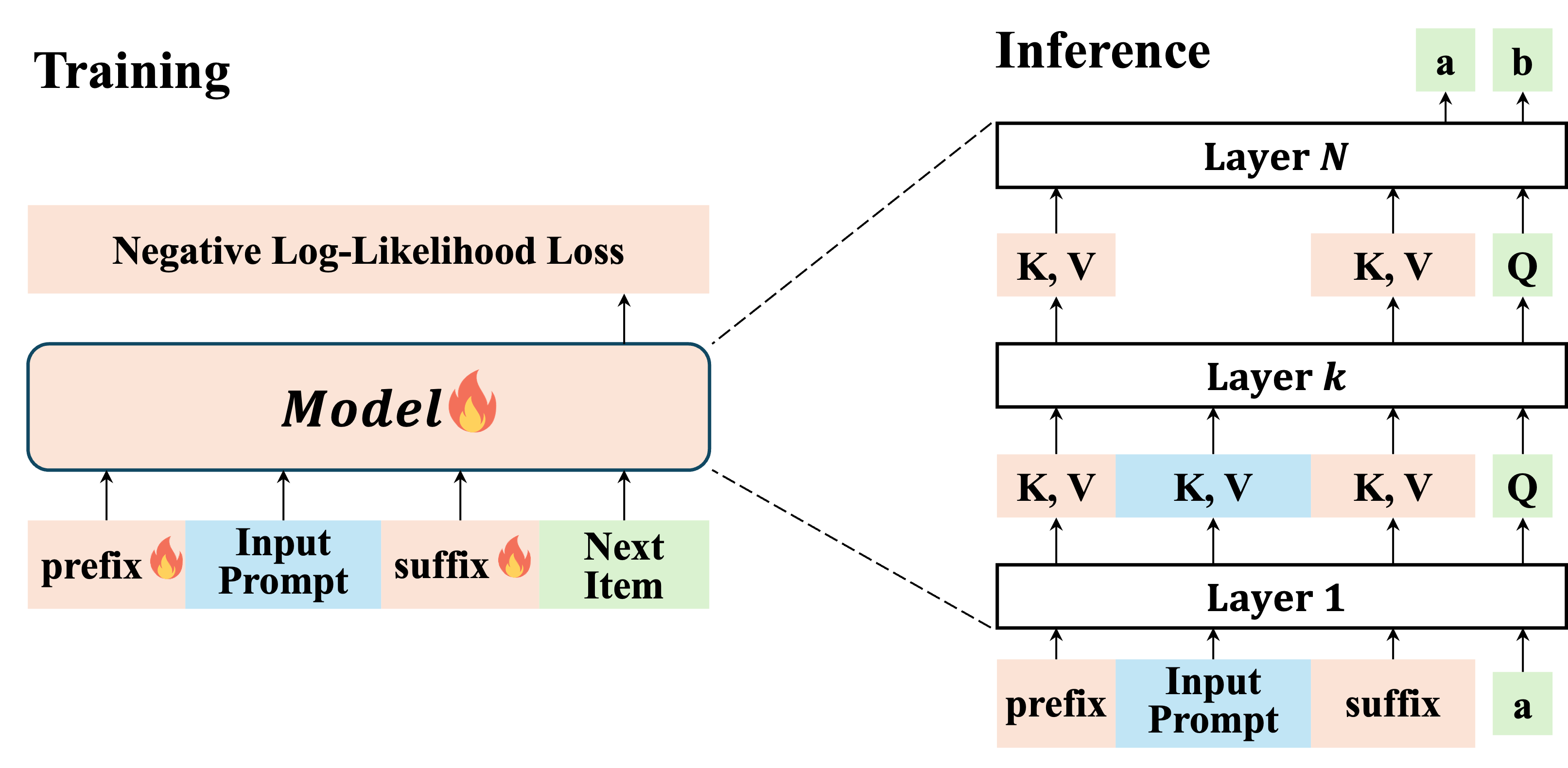}
    \caption{Overview of the proposed EARN. During training, 
    the prefix register and the suffix register are also trainable. During inference, after layer $\bm k$, EARN removes the prompt tokens to achieve acceleration.}
    \label{fig:method}
\end{figure}

In order to improve the inference efficiency while ensuring the recommendation effectiveness, we proposed \textbf{EARN}, which achieves inference acceleration through register tokens.
The overview of our method is presented in Figure~\ref{fig:method}.

\subsection{Register Tokens}

\sss{Prefix Register}
We introduce the \textit{prefix register}, \ie a set of learnable virtual tokens placed at the beginning of the input prompt. These tokens are designed to learn task-specific instructions, effectively signaling to the LLM that the current task is a recommendation problem.
This method is inspired by the head sinks of the dual attention sinks phenomenon as elaborated in Section~\ref{sec:sink}. Head sinks suggest that the head tokens can effectively divert attention away from task-irrelevant tokens in the subsequent layers. 
Drawing on the practice of prompt tuning~\cite{liu2024gpt, liu2021p}, where learnable virtual tokens are used to represent task instructions, we replace the BOS token, which the attention always sinks into, with the learnable prefix register. This substitution not only fulfills the attention-diverting function of the BOS token, but also serves the role of task indication to tell the model that this is a recommendation task.

\sss{Suffix Register}
We introduce the \textit{suffix register}, \ie a set of learnable virtual tokens placed at the end of the input prompt, designed to summarize historical interactions and extract key information. 
This method is inspired by the tail sinks of the dual attention sinks phenomenon as elaborated in Section~\ref{sec:sink}. 
The final tokens in a sequence have visibility over all preceding tokens, endowing them with summarization capabilities. Prior studies have observed that semantically meaningless tokens can effectively summarize preceding information during LLM inference~\cite{chen2024sepllm,pang2024anchor}. The occurrence of the tail sink phenomenon in the LLMRec task further corroborates this point. Building on this insight, we place learnable virtual tokens at the end of the input prompt and leverage the first $k$ layers of the LLM to amplify this summarization capability.

\subsection{Overall Pipeline}

\sss{Training}
During training, we employ next token prediction to train our model, computing the loss on the target item identifier, \ie minimizing the negative log-likelihood loss between the predicted next-item probabilities and the ground truth items. Formally, given a chronological sequence of user interactions $H=[i_1, i_2, ..., i_{T+1}]$, we construct the input sequence as:
\begin{equation}
    X = [R_{\text{prefix}}; P\text{rompt}; R_{\text{suffix}}] \ ,
\end{equation}
\begin{equation}
    Y = [i_{T+1}] \ ,
\end{equation}
where $R_{\text{prefix}}$ and $R_{\text{suffix}}$ denote the prefix register and the suffix register, respectively. $P\text{rompt}$ is the user historical interactions $[i_1, i_2, ..., i_{T}]$ with the task instruction, and $Y$ is the target item identifier $i_{T+1}$. The loss function is formulated as,
\begin{equation}\small
    \mathcal{L} = -\sum_{j=1}^{|Y|} \log P(Y_{j} \mid X, Y_{<j}) \ ,
\end{equation}
where $X$ is the input sequence of the sample, $Y$ is the corresponding item identifier, $Y_{j}$ is the $j$-th token in the item identifier $Y$, and $Y_{<j}$ denotes the tokens preceding $Y_{j}$ in the item identifier. 

When calculating the loss function, the main difference between our method and ordinary finetuning is the calculation of attention, where we exclude prompt tokens after layer $k$, \ie
\begin{itemize}[leftmargin=*]
    \item For layer $l \leq k$: All tokens participate in the attention calculation.
    \item For layer $l > k$: Only register and generated tokens participate in the attention calculation.
\end{itemize}

\sss{Inference}
During inference, for a model with $N$ layers, the generation of each token consists of two processes:
\begin{enumerate}[leftmargin=*]
    \item \textbf{Full Computation Process (First $k$ Layers)}: All tokens including task instructions, user historical interactions, prefix register and suffix register participate in the model computation. The model processes the complete input sequence to establish rich contextual representations.
    
    \item \textbf{Register-Focused Process (Remaining $N-k$ Layers)}: After $k$ layers, we remove the original prompt tokens (task instructions and historical interactions), retaining only the register tokens and newly generated tokens. The attention mechanism in subsequent layers only operates on this reduced set of tokens:
    
    \begin{itemize}[leftmargin=*]
        \item In the \textit{prefilling} stage, the input hidden state of layer $k$ is transformed from $\bm{X}=[\bm{X}_{\text{prefix}}; \bm{X}_{\text{Prompt}}; \bm{X}_{\text{suffix}}]$ to $\bm{X'}=[\bm{X}_{\text{prefix}}; \bm{X}_{\text{suffix}}]$, significantly reducing the computational load and the initial KV Cache size.
        \item In the \textit{decoding} stage, the input hidden state of layer $k$ is transformed from $\bm{X}=[\bm{X}_{\text{prefix}}; \bm{X}_{\text{Prompt}}; \bm{X}_{\text{suffix}}; \bm{X}_{\text{generated}}]$ to $\bm{X'}=[\bm{X}_{\text{prefix}}; \bm{X}_{\text{suffix}}; \bm{X}_{\text{generated}}]$, significantly reducing KV Cache that needs to be accessed during \textit{decoding}.
    \end{itemize}
\end{enumerate}

This approach maintains the model's ability to leverage historical information through the learned register representations while avoiding redundant computations on lengthy prompt tokens.

\subsection{Efficiency Analysis}
Due to the removal of prompt tokens in the subsequent layers, we can significantly reduce both the computational load and memory footprint. 
Assume the model has $N$ layers, $n_h$ heads, an attention dimension of $d_a$, a hidden state dimension of $d_h$, and an intermediate dimension of $d_f$. The length of the input prompt is $L$, and the number of register tokens is $r \ll L$. 
If we remove the prompt tokens after $k$ layers, the efficiency promotion can be analyzed through three key aspects:

\begin{itemize}[leftmargin=*]
    \item \textbf{Computation Complexity}: 
    The FLOPs of the vanilla LLM are $N\ FLOPs_{LLM\ layer} = 4NL[n_h d_a(2d_h + L)+d_h d_f]$ (see Appendix~\ref{app:inference_speed} for detailed derivation).
    In our EARN, it becomes $k\ FLOPs_{LLM\ layer} + (N-k)\ FLOPs_{Register\ layer}$. Thus, the attention complexity ratio $\gamma_{\text{attn}}$ becomes:
    \begin{equation}
    \begin{aligned}
    \Gamma_{attn} &= \frac{k\ FLOPs_{LLM\ layer} + (N-k)\ FLOPs_{Register\ layer}}{N\ FLOPs_{LLM\ layer}} \\
    &= \frac{k}{N} + (1-\frac{k}{N})\frac{r[n_h d_a(2d_h + r)+d_h d_f]}{L[n_h d_a(2d_h + L)+d_h d_f]} \\
    &\approx \frac{k}{N}. 
    \end{aligned}
    \ 
    \end{equation}

    \item \textbf{KV Cache Size}: The memory size of the KV Cache is $M_{KV} N_{KV}$, where $M_{KV}$ represents the memory size of each KV pair, and $N_{KV}$ denotes the total number of KV pairs. The original $N_{KV}$ would be $n_h NL\ ,$ whereas now it becomes $n_h(kL + (N - k)r)\ .$ Specifically, the KV Cache can be shrunk to $\frac{kL+(N-k)r}{NL}$ of its original size, reduced by $\frac{(N-k)(L-r)}{NL}.$ That is to say,
    the KV cache size reduction ratio $\gamma_{\text{cache}}$ is:
    \begin{equation}
    \begin{aligned}
    \Gamma_{cache} &= 1 - \frac{kL + (N - k)r}{NL} \\ 
    &= \frac{(N-k)(L-r)}{NL} \\
    &\approx \frac{N-k}{N}. 
    \end{aligned}
    \ 
    \end{equation}

    \item \textbf{Theoretical Speedup}: Assuming that FLOPS of the device is $v_c$, and HBM rate is $v_m$. The estimated time cost is 
    \begin{equation}
        T=T_P+T_D=T_P+n_{generate}T_d \ ,
    \end{equation}
    where 
    \begin{equation}
        T_P=\frac{FLOPs_{prefilling}}{v_c} \ ,
    \end{equation}
    \begin{equation}
        T_d=max(\frac{Cache}{v_m},\frac{FLOPs_{attn}}{v_c})+\frac{FLOPs_{FFN}}{v_c} \ .
    \end{equation}
    The theoretical speedup is
    \begin{equation}
        \Omega = \frac{T_{vanilla}}{T_{EARN}} \approx \frac{N}{k} \ .
    \end{equation}
\end{itemize}

For typical values ($N=32,\ n_h=32,\ d_a=128,\ d_h=4096, d_f=11008,\ L=512,\ k=8,\ r=2$), EARN reduces the KV Cache size to 75\% of the original, and the speedup ratio can reach 4x.

\section{Experiment}
In this section, we conduct extensive experiments to answer the following research questions: 
\begin{itemize}[leftmargin=*]
    \item \textbf{RQ1}: How does our proposed EARN perform compared to common inference acceleration methods? 
    \item \textbf{RQ2}: How is the efficiency scalability of EARN under different batch sizes and sequence lengths?
    \item \textbf{RQ3}: How do different hyper-parameters affect the trade-off between inference efficiency and recommendation effectiveness of EARN?
    \item \textbf{RQ4}: How do different components contribute to EARN?
\end{itemize}

\subsection{Experimental Settings}

\sss{Models and Datasets}
We conduct experiments on two mainstream LLMRec methods: LC-Rec~\cite{zheng2024adapting} and TIGER~\cite{rajput2023recommender}, 
with two distinct LLM architectures: Llama-7B~\cite{touvron2023llama} using multi-head attention and Qwen2.5-7B~\cite{qwen2.5} using grouped-query attention. 
We test EARN on three real-world recommendation datasets: 1) \textbf{Beauty} contains user interactions with the beauty products. 2) \textbf{Games} covers user interactions with the video games. 3) \textbf{MovieLens-1M} collects user interactions with movies.
More details of the datasets and experimental implementation details can be found in Appendix~\ref {app:dataset}. 

\sss{Baselines}
We compare our approach against commonly used practices and existing SOTA methods, serving as our baselines:
\begin{itemize}[leftmargin=*]
\item \textbf{Basic Methods}
    \begin{itemize}[leftmargin=*]
    \item \textit{Finetune}: Standard full finetuning of the model.
    \item \textit{SkipLayers}: Skipping redundant subsequent layers.
    \end{itemize}
\item \textbf{Prompt Compression}: Methods targeting the \textit{prefilling} stage.
    \begin{itemize}[leftmargin=*]
    \item \textit{POD}~\cite{li2023prompt}: Distilling task instructions in the prompt into few virtual tokens.
    \item \textit{500xCompressor}~\cite{li2024500xcompressor}: Compressing the context (user historical interactions in our scenario) within the prompt into one single special token, by employing frozen original LLM and trainable additional LoRA parameters.
    \end{itemize}
\item \textbf{Cache Compression}: Methods targeting the \textit{decoding} stage.
    \begin{itemize}[leftmargin=*]
    \item \textit{StreamingLLM}~\cite{xiao2023efficient}: Statically retaining initial tokens and fixed-length recent tokens.
    \item \textit{SnapKV}~\cite{li2024snapkv}: Dynamically caching clustered important tokens.
    \end{itemize}
\item \textbf{Gist Methods}: Methods using the register token idea.
    \begin{itemize}[leftmargin=*]
    \item \textit{Gist}~\cite{mu2023learning}: Employing LLM itself to compress task instructions in the prompt into few gist tokens.
    \item \textit{AnLLM}~\cite{pang2024anchor}: Employing LLM itself to compress segmented sentences into the anchor token at the end of the sentence.
    \end{itemize}
\end{itemize}

\sss{Evaluation Metrics}
Referring to previous work in the field of recommendation systems and LLM inference acceleration[8], we use the following metrics to evaluate our method:

\begin{itemize}[leftmargin=*]
\item \textbf{Time Efficiency}
    \begin{itemize}[leftmargin=*]
    \item \textit{Walltime Speedup} $\omega$: The actual test speedup relative to vanilla auto-regressive decoding by comparing the wall clock time.
    \item \textit{Throughput} $\tau$: The number of new tokens that the model generates per second.
    \end{itemize}
\item \textbf{Space Efficiency}
    \begin{itemize}[leftmargin=*]
    \item \textit{KV Cache Reduction} $\gamma$: The empirically measured percentage reduction of the KV Cache.
    \item \textit{KV Cache Memory} $\sigma$: The average GPU memory usage (GB) of the KV Cache when generating the last token.
    \end{itemize}
\item \textbf{Recommendation Effectiveness}
    \begin{itemize}[leftmargin=*]
    \item \textit{Recall@\textit{K}} (\textit{R@K}): A widely used measure of the model's ability to retrieve relevant items, which is defined as the proportion of relevant items that are recommended out of the total number of relevant items available.
    \item \textit{NDCG@\textit{K}} (\textit{N@K}): Normalized Discounted Cumulative Gain (NDCG) is a measure that takes into account both the order of relevant items
    and the relevance score of each item.
    \end{itemize}
\end{itemize}

\begin{table*}[!ht]
    \centering
    \setlength{\abovecaptionskip}{0cm}
    \setlength{\belowcaptionskip}{0cm}
    \caption{Overall performance comparison between the baselines and EARN instantiated on LC-Rec. The best results are highlighted in bold and the second-best results are underlined.}
    \label{tab:overall}
    \fontsize{8pt}{8.8pt}\selectfont
    \begin{tabular}{c|c|l|rr|rr|ccccc}
    \toprule
    \multirow{2}{*}{\textbf{Dataset}} & \multirow{2}{*}{\textbf{Model}} & \multirow{2}{*}{\textbf{Method}} & \multicolumn{2}{c|}{\textbf{Time Efficiency}} & \multicolumn{2}{c|}{\textbf{Space Efficiency}} & \multicolumn{4}{c}{\textbf{Recommendation Effectiveness}} \\
    &  &  & $\bm\omega$ & $\bm\tau$ & $\bm\gamma$ & $\bm\sigma$ & \textbf{R@10} & \textbf{R@20} & \textbf{N@10} & \textbf{N@20} \\
    \midrule
    \multirow{18}{*}{\textbf{Beauty}} & \multirow{9}{*}{\textbf{Llama}} & \textbf{Finetune} & 1.00 & 505.2 & 0.0 & 85.45 & \uline{0.0145} & \uline{0.0225} & \uline{0.0084} & \uline{0.0108} \\ 
    ~ & ~ & \textbf{SkipLayers} & 1.79 & 895.4 & 44.4 & 47.50 & 0.0013 & 0.0013 & 0.0013 & 0.0013 \\ 
    ~ & ~ & \textbf{POD} & 1.15 & 585.0 & 14.7 & 72.87 & 0.0045 & 0.0074 & 0.0032 & 0.0041 \\ 
    ~ & ~ & \textbf{500xCompressor} & \uline{2.31} & \uline{1168.6} & 74.8 & 21.55 & 0.0005 & 0.0006 & 0.0002 & 0.0003 \\ 
    ~ & ~ & \textbf{StreamingLLM} & 1.22 & 611.2 & \textbf{96.4} & \textbf{3.09} & 0.0005 & 0.0005 & 0.0004 & 0.0004 \\ 
    ~ & ~ & \textbf{SnapKV} & 1.20 & 600.7 & \uline{94.5} & \uline{4.73} & 0.0054 & 0.0061 & 0.0030 & 0.0032 \\ 
    ~ & ~ & \textbf{Gist} & 1.18 & 597.6 & 17.5 & 70.50 & 0.0048 & 0.0077 & 0.0028 & 0.0036 \\
    ~ & ~ & \textbf{AnLLM} & 1.24 & 625.1 & 92.2 & 6.70 & 0.0000 & 0.0000 & 0.0000 & 0.0000 \\
    ~ & ~ & \cellcolor{gray!16}\textbf{EARN} & \cellcolor{gray!16}\textbf{3.79} & \cellcolor{gray!16}\textbf{1844.8} & \cellcolor{gray!16}80.5 & \cellcolor{gray!16}16.68 & \cellcolor{gray!16}\textbf{0.0167} & \cellcolor{gray!16}\textbf{0.0265} & \cellcolor{gray!16}\textbf{0.0095} & \cellcolor{gray!16}\textbf{0.0124} \\
    \cline{2-11} \rule{0pt}{1em}
    ~ & \multirow{9}{*}{\textbf{Qwen}} & \textbf{Finetune} & 1.00 & 622.1 & 0.0 & 14.08 & \uline{0.0145} & \uline{0.0248} & \uline{0.0087} & \uline{0.0117} \\ 
    ~ & ~ & \textbf{SkipLayers} & 1.73 & 1056.1 & 58.0 & 5.92 & 0.0000 & 0.0000 & 0.0000 & 0.0000 \\ 
    ~ & ~ & \textbf{POD} & 1.09 & 679.3 & 8.4 & 12.90 & 0.0082 & 0.0127 & 0.0047 & 0.0061 \\ 
    ~ & ~ & \textbf{500xCompressor} & \uline{2.56} & \uline{1587.1} & \uline{91.1} & \uline{1.26} & 0.0003 & 0.0003 & 0.0001 & 0.0001 \\ 
    ~ & ~ & \textbf{StreamingLLM} & 1.05 & 652.6 & \textbf{92.0} & \textbf{1.12} & 0.0088 & 0.0147 & 0.0058 & 0.0075 \\ 
    ~ & ~ & \textbf{SnapKV} & 1.02 & 634.5 & 69.5 & 4.29 & 0.0097 & 0.0165 & 0.0058 & 0.0077 \\ 
    ~ & ~ & \textbf{Gist} & 1.15 & 715.4 & 20.0 & 11.30 & 0.0084 & 0.0161 & 0.0050 & 0.0074 \\
    ~ & ~ & \textbf{AnLLM} & 1.06 & 659.3 & 80.5 & 2.70 & 0.0000 & 0.0000 & 0.0000 & 0.0000 \\
    ~ & ~ & \cellcolor{gray!16}\textbf{EARN} & \cellcolor{gray!16}\textbf{2.71} & \cellcolor{gray!16}\textbf{1662.6} & \cellcolor{gray!16}75.2 & \cellcolor{gray!16}3.49 & \cellcolor{gray!16}\textbf{0.0155} & \cellcolor{gray!16}\textbf{0.0265} & \cellcolor{gray!16}\textbf{0.0091} & \cellcolor{gray!16}\textbf{0.0122} \\ 
    \hline \rule{0pt}{1em}
    \multirow{18}{*}{\textbf{Games}} & \multirow{9}{*}{\textbf{Llama}} & \textbf{Finetune} & 1.00 & 571.4 & 0.0 & 78.10 & \uline{0.0167} & \uline{0.0273} & \uline{0.0106} & \uline{0.0138} \\ 
    ~ & ~ & \textbf{SkipLayers} & 1.85 & 1045.3 & 45.0 & 42.93 & 0.0002 & 0.0002 & 0.0003 & 0.0003 \\ 
    ~ & ~ & \textbf{POD} & 1.09 & 623.3 & 15.9 & 65.69 & 0.0088 & 0.0147 & 0.0052 & 0.0070 \\ 
    ~ & ~ & \textbf{500xCompressor} & \uline{2.11} & \uline{1205.3} & 72.5 & 21.45 & 0.0014 & 0.0021 & 0.0009 & 0.0011 \\ 
    ~ & ~ & \textbf{StreamingLLM} & 1.23 & 702.5 & \textbf{96.0} & \textbf{3.13} & 0.0013 & 0.0013 & 0.0014 & 0.0014 \\ 
    ~ & ~ & \textbf{SnapKV} & 1.22 & 688.5 & \uline{93.5} & \uline{5.04} & 0.0102 & 0.0110 & 0.0070 & 0.0072 \\ 
    ~ & ~ & \textbf{Gist} & 1.12 & 639.9 & 18.0 & 64.00 & 0.0091 & 0.0146 & 0.0053 & 0.0070 \\
    ~ & ~ & \textbf{AnLLM} & 1.25 & 715.5 & 91.6 & 6.60 & 0.0003 & 0.0003 & 0.0003 & 0.0003 \\
    ~ & ~ & \cellcolor{gray!16}\textbf{EARN} & \cellcolor{gray!16}\textbf{3.53} & \cellcolor{gray!16}\textbf{1930.6} & \cellcolor{gray!16}80.8 & \cellcolor{gray!16}14.98 & \cellcolor{gray!16}\textbf{0.0180} & \cellcolor{gray!16}\textbf{0.0291} & \cellcolor{gray!16}\textbf{0.0107} & \cellcolor{gray!16}\textbf{0.0142} \\ 
    \cline{2-11} \rule{0pt}{1em}
    ~ & \multirow{9}{*}{\textbf{Qwen}} & \textbf{Finetune} & 1.00 & 568.1 & 0.0 & 13.03 & \uline{0.0193} & \textbf{0.0316} & \textbf{0.0127} & \uline{0.0155} \\ 
    ~ & ~ & \textbf{SkipLayers} & 1.52 & 858.4 & 57.3 & 5.57 & 0.0000 & 0.0000 & 0.0000 & 0.0000 \\ 
    ~ & ~ & \textbf{POD} & 1.35 & 767.5 & 8.1 & 11.98 & 0.0129 & 0.0209 & 0.0075 & 0.0100 \\ 
    ~ & ~ & \textbf{500xCompressor} & \uline{2.60} & \uline{1475.4} & \uline{97.1} & \uline{0.38} & 0.0013 & 0.0023 & 0.0008 & 0.0011 \\ 
    ~ & ~ & \textbf{StreamingLLM} & 1.05 & 594.7 & \textbf{98.4} & \textbf{0.21} & 0.0129 & 0.0202 & 0.0075 & 0.0098 \\ 
    ~ & ~ & \textbf{SnapKV} & 1.02 & 576.2 & 67.9 & 4.19 & 0.0138 & 0.0226 & 0.0083 & 0.0110 \\ 
    ~ & ~ & \textbf{Gist} & 1.41 & 801.3 & 22.9 & 10.00 & 0.0121 & 0.0154 & 0.0077 & 0.0088 \\
    ~ & ~ & \textbf{AnLLM} & 1.09 & 616.9 & 95.2 & 0.60 & 0.0003 & 0.0003 & 0.0003 & 0.0003 \\
    ~ & ~ & \cellcolor{gray!16}\textbf{EARN} & \cellcolor{gray!16}\textbf{3.11} & \cellcolor{gray!16}\textbf{1711.3} & \cellcolor{gray!16}75.1 & \cellcolor{gray!16}3.24 & \cellcolor{gray!16}\textbf{0.0197} & \cellcolor{gray!16}\uline{0.0312} & \cellcolor{gray!16}\uline{0.0122} & \cellcolor{gray!16}\textbf{0.0157} \\ 
    \hline \rule{0pt}{1em}
    \multirow{18}{*}{\textbf{MovieLens}} & \multirow{9}{*}{\textbf{Llama}} & \textbf{Finetune} & 1.00 & 704.3 & 0.0 & 55.39 & \uline{0.0247} & \uline{0.0449} & \uline{0.0197} & \uline{0.0288} \\ 
    ~ & ~ & \textbf{\textbf{SkipLayers}} & 2.52 & 1302.9 & 67.6 & 17.93 & 0.0022 & 0.0022 & 0.0043 & 0.0043 \\ 
    ~ & ~ & \textbf{POD} & 1.17 & 827.4 & 18.0 & 45.40 & 0.0066 & 0.0118 & 0.0062 & 0.0087 \\ 
    ~ & ~ & \textbf{500xCompressor} & \uline{1.92} & \uline{1352.8} & 61.1 & 21.54 & 0.0004 & 0.0005 & 0.0003 & 0.0003 \\ 
    ~ & ~ & \textbf{StreamingLLM} & 1.20 & 836.2 & \textbf{95.2} & \textbf{2.66} & 0.0000 & 0.0000 & 0.0000 & 0.0000 \\ 
    ~ & ~ & \textbf{SnapKV} & 1.19 & 829.2 & \uline{91.9} & \uline{4.50} & 0.0000 & 0.0000 & 0.0000 & 0.0000 \\ 
    ~ & ~ & \textbf{Gist} & 1.01 & 711.3 & 22.3 & 43.10 & 0.0232 & 0.0421 & 0.0118 & 0.0250 \\
    ~ & ~ & \textbf{AnLLM} & 1.20 & 846.9 & 89.5 & 5.80 & 0.0006 & 0.0008 & 0.0005 & 0.0006 \\
    ~ & ~ & \cellcolor{gray!16}\textbf{EARN} & \cellcolor{gray!16}\textbf{3.21} & \cellcolor{gray!16}\textbf{2250.2} & \cellcolor{gray!16}79.7 & \cellcolor{gray!16}11.26 & \cellcolor{gray!16}\textbf{0.0259} & \cellcolor{gray!16}\textbf{0.0452} & \cellcolor{gray!16}\textbf{0.0247} & \cellcolor{gray!16}\textbf{0.0341} \\ 
    \cline{2-11} \rule{0pt}{1em}
    ~ & \multirow{9}{*}{\textbf{Qwen}} & \textbf{Finetune} & 1.00 & 752.9 & 0.0 & 10.71 & \uline{0.0289} & \uline{0.0421} & \uline{0.0247} & \uline{0.0315} \\ 
    ~ & ~ & \textbf{SkipLayers} & 1.87 & 1124.7 & 76.6 & 2.50 & 0.0000 & 0.0000 & 0.0000 & 0.0000 \\ 
    ~ & ~ & \textbf{POD} & 1.41 & 1061.7 & 40.1 & 6.42 & 0.0139 & 0.0177 & 0.0115 & 0.0131 \\ 
    ~ & ~ & \textbf{500xCompressor} & \uline{2.09} & \uline{1577.8} & \uline{77.6} & \uline{2.40} & 0.0006 & 0.0009 & 0.0006 & 0.0007 \\ 
    ~ & ~ & \textbf{StreamingLLM} & 1.05 & 766.8 & \textbf{88.9} & \textbf{1.19} & 0.0258 & 0.0372 & 0.0197 & 0.0244 \\ 
    ~ & ~ & \textbf{SnapKV} & 1.00 & 731.9 & 76.5 & 2.52 & 0.0248 & 0.0428 & 0.0192 & 0.0266 \\ 
    ~ & ~ & \textbf{Gist} & 1.06 & 799.0 & 30.0 & 7.50 & 0.0287 & 0.0441 & 0.0244 & 0.0305 \\
    ~ & ~ & \textbf{AnLLM} & 1.03 & 772.4 & 75.9 & 2.60 & 0.0022 & 0.0022 & 0.0043 & 0.0043 \\
    ~ & ~ & \cellcolor{gray!16}\textbf{EARN} & \cellcolor{gray!16}\textbf{2.84} & \cellcolor{gray!16}\textbf{2136.7} & \cellcolor{gray!16}66.7 & \cellcolor{gray!16}3.56 & \cellcolor{gray!16}\textbf{0.0298} & \cellcolor{gray!16}\textbf{0.0591} & \cellcolor{gray!16}\textbf{0.0252} & \cellcolor{gray!16}\textbf{0.0382} \\ 
    \bottomrule
    \end{tabular}
\end{table*}

\subsection{Overall Performance (RQ1)}
The overall results of baselines and EARN instantiated on LC-Rec on three datasets and two LLM models are presented in Table~\ref{tab:overall}. The results instantiated on TIGER are similar, which are moved to Appendix~\ref{app:TIGER}. We draw a few observations as follows:

\begin{itemize}[leftmargin=*]
    \item Qwen-based implementations outperform Llama-based implementations in both inference efficiency and recommendation effectiveness. This superiority arises from: \textbf{1)} Qwen's grouped-head attention architecture achieves a smaller KV Cache and lower inference latency while maintaining semantic capability; \textbf{2)} Qwen's expanded tokenizer vocabulary (151,851 vs. Llama's 32,000) that shortens input prompts, reducing semantic fragmentation; \textbf{3)} Qwen is trained on a more massive dataset with approximately 18 trillion tokens, which provides it with a richer knowledge base and better language understanding capabilities. 
    \item Among all inference acceleration baselines, although cache compression methods (StreamingLLM and SnapKV) achieve the highest cache reduction (over 90\%), their end-to-end speedup is not as significant as prompt compression methods. This is because cache compression methods do not optimize the \textit{prefilling} stage, whereas prompt compression methods (POD and 500xCompressor) reduce the input sequence length in the \textit{prefilling} stage, thereby improving both computational and memory efficiency. This highlights the greater potential for acceleration in the \textit{prefilling} stage for LLMRec. However, despite the notable inference efficiency gains of the baselines, they suffer from catastrophic recommendation effectiveness degradation. This highlights that user historical interactions in recommendation scenarios cannot be simply compressed at a high ratio using NLP methods. 
    \item Although gist methods (Gist and AnLLM) also employ the idea of register tokens, they fail to achieve a significant speedup. This is because they utilize all layers of the LLM to compress information, resulting in substantial computational costs. In contrast, EARN only uses the first \( k \) layers, significantly reducing the computational burden. In terms of space efficiency, Gist only compresses the task instruction within the input sequence, thus saving only about 20\% of cache memory. Although AnLLM can reduce cache memory by 90\%, it comes at the cost of catastrophic loss in recommendation effectiveness. AnLLM's recommendation effectiveness is nearly zero. This is due to its attempt to compress the intricate user interaction history into a single token, which is highly challenging and leads to severe information loss and poor performance. Gist shows a decline in recommendation effectiveness. This is because Gist compresses the task instruction in isolation within the language space, while the items in recommendations are not within the LLM's language space. Without direct interaction between the task instruction and historical items, the LLM struggles to understand the recommendation task's requirements to predict items outside its vocabulary. This impedes the method's applicability to recommendation tasks.
    \item EARN significantly achieves SOTA time efficiency (2.71-3.79x speedup) with excellent space efficiency (66.7-80.8\% cache reduction), and also demonstrates a notable improvement over Finetune in terms of recommendation effectiveness. This suggests that the register tokens used in EARN can effectively summarize useful information, thereby ensuring that recommendation effectiveness is maintained while inference efficiency is improved.
\end{itemize}

\subsection{Efficiency Scalability (RQ2)}
To investigate the efficiency scalability of EARN, we evaluate the inference efficiency of EARN as compared to Finetune under varying batch sizes and sequence lengths.

\sss{Efficiency under Different Batch Sizes}
\begin{figure}[t]
    \vspace{-0.2cm}
    \setlength{\abovecaptionskip}{0cm}
    \setlength{\belowcaptionskip}{0cm}
    \centering
    \includegraphics[width=1\linewidth]{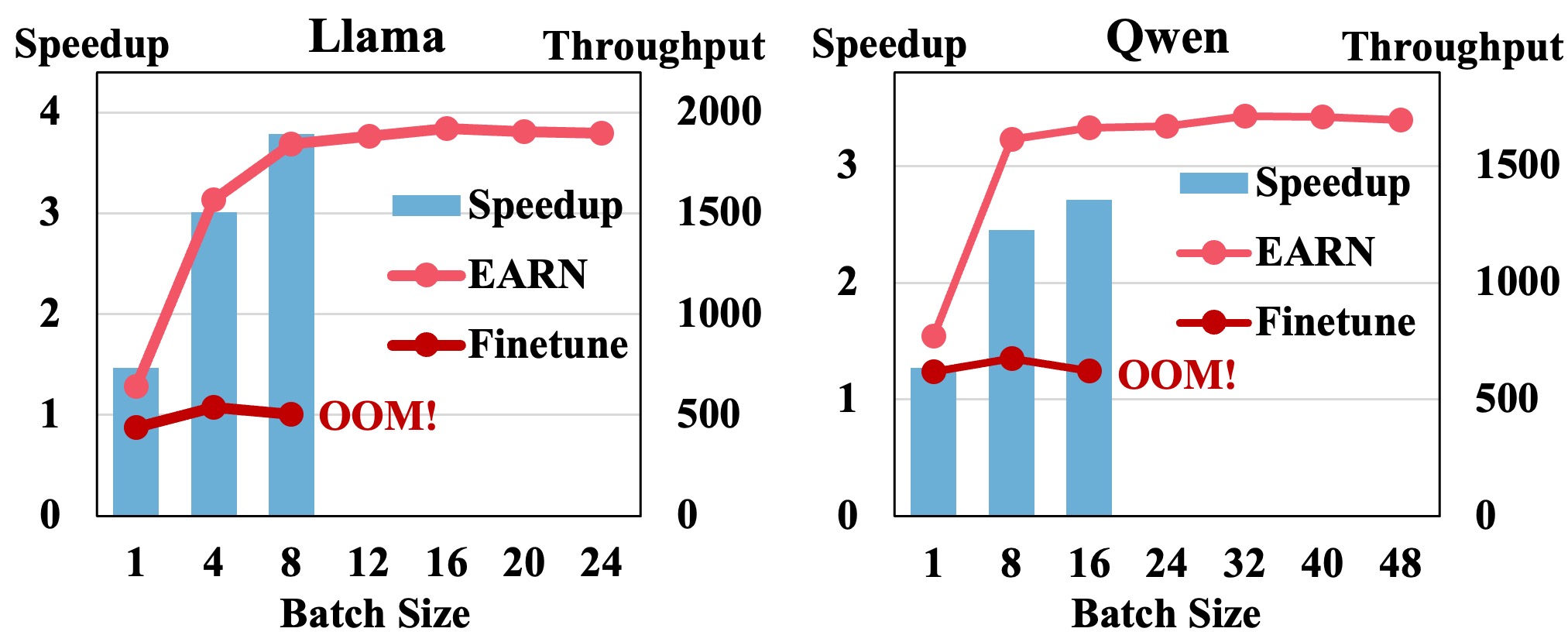}
    \caption{Efficiency under different batch sizes.}
    \label{fig:bs-throughput}
\end{figure}
\begin{figure}[t]
    \vspace{-0.2cm}
    \setlength{\abovecaptionskip}{0cm}
    \setlength{\belowcaptionskip}{0cm}
    \centering
    \includegraphics[width=1\linewidth]{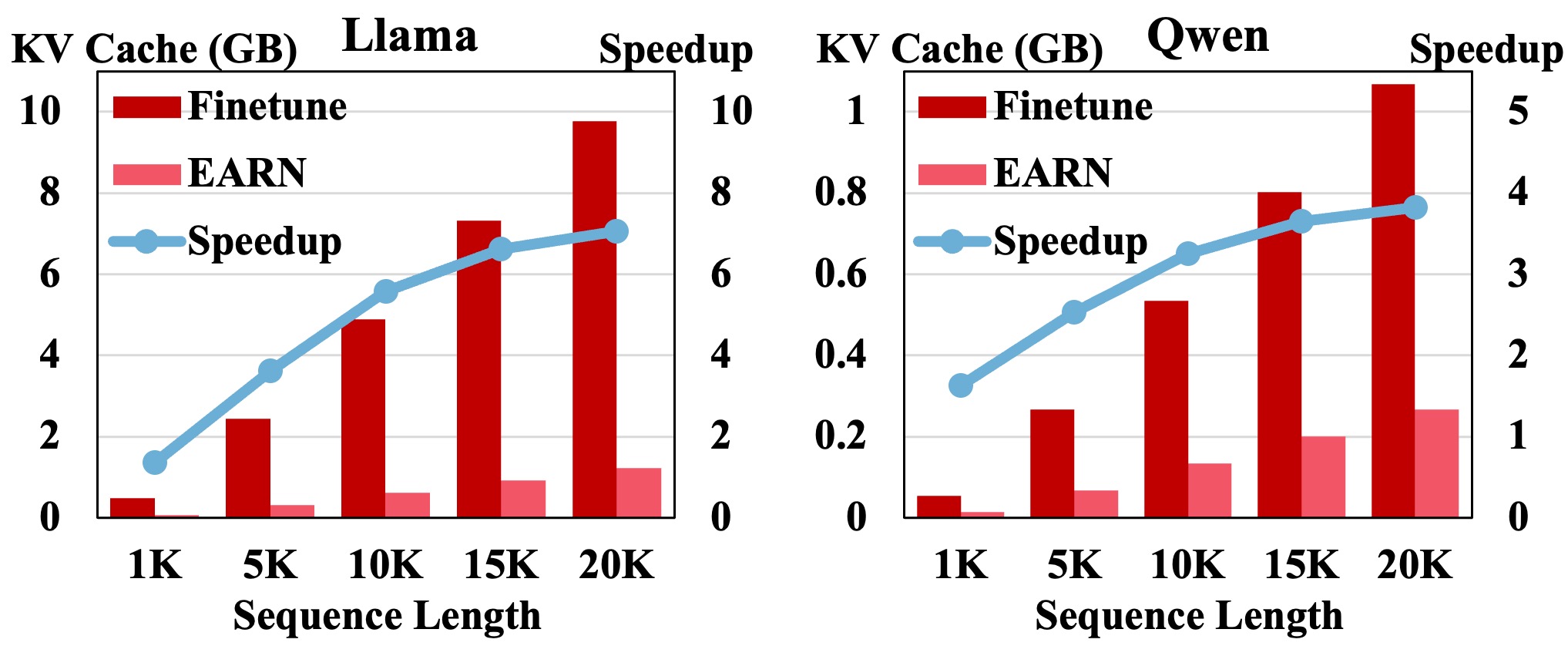}
    \caption{Efficiency under different sequence lengths.}
    \label{fig:len-KV}
\end{figure}

To verify the acceleration effect of EARN under different batch sizes, we test the inference efficiency of EARN and Finetune with batch size in \{1, 4, 8, · · · , 24\} on Llama and \{1, 8, 16, · · · , 48\} on Qwen.
As shown in Figure~\ref{fig:bs-throughput}, EARN maintains significantly higher throughput than Finetune across all batch sizes. The speedup of EARN over Finetune increases as the batch size grows. Moreover, EARN can handle a larger batch size, which is particularly useful for practical scenarios. For example, on Llama, Finetune runs into out-of-memory (OOM) errors when the batch size exceeds 8, while EARN can still support larger batch sizes. The results clearly demonstrate that EARN significantly improves efficiency compared to Finetune across different batch sizes. Its ability to maintain high throughput on large batch sizes highlights its superior efficiency and practical inference capabilities. This makes EARN a more suitable choice for real-world applications where efficient and scalable inference is crucial.

\sss{Efficiency under Different Sequence Lengths}

In industrial settings, user interaction histories can be extensive. Therefore, it is essential to assess the efficiency of EARN on longer sequences. To simulate this scenario, we pad our dataset to specific lengths (1K, 5K, · · ·, 20K) to test the acceleration effects.
From Figure~\ref{fig:len-KV}, we can see EARN consistently uses significantly less KV Cache than Finetune across all sequence lengths. Moreover, the larger the sequence length, the more significant the acceleration effect, achieving a speedup of up to 7x when the sequence length is 20K on Llama.
This is highly desirable in practice, especially for long-term sequential recommendation, where user historical interaction sequences are stored for a long time, thus learning user preferences comprehensively.
The results clearly demonstrate that EARN significantly improves efficiency compared to Finetune across different sequence lengths. Its ability to maintain a low KV Cache size and achieve higher speedup, especially for longer sequences, highlights its superior efficiency and practical inference capabilities. This makes EARN a more suitable choice for real-world applications where efficient and scalable inference is crucial, particularly when dealing with long sequences.

\subsection{Hyper-parameter Analysis (RQ3)}
To investigate the trade-off between inference efficiency and recommendation effectiveness caused by variations in hyper-parameters, we train and evaluate EARN with different register layer depth $k$ and register token number $n$.
\sss{Effect of Register Layer Depth $\bm k$}

\begin{figure}[t]
\vspace{-0.2cm}
    \centering
    \setlength{\abovecaptionskip}{0cm}
    \setlength{\belowcaptionskip}{0cm}
    \includegraphics[width=1\linewidth]{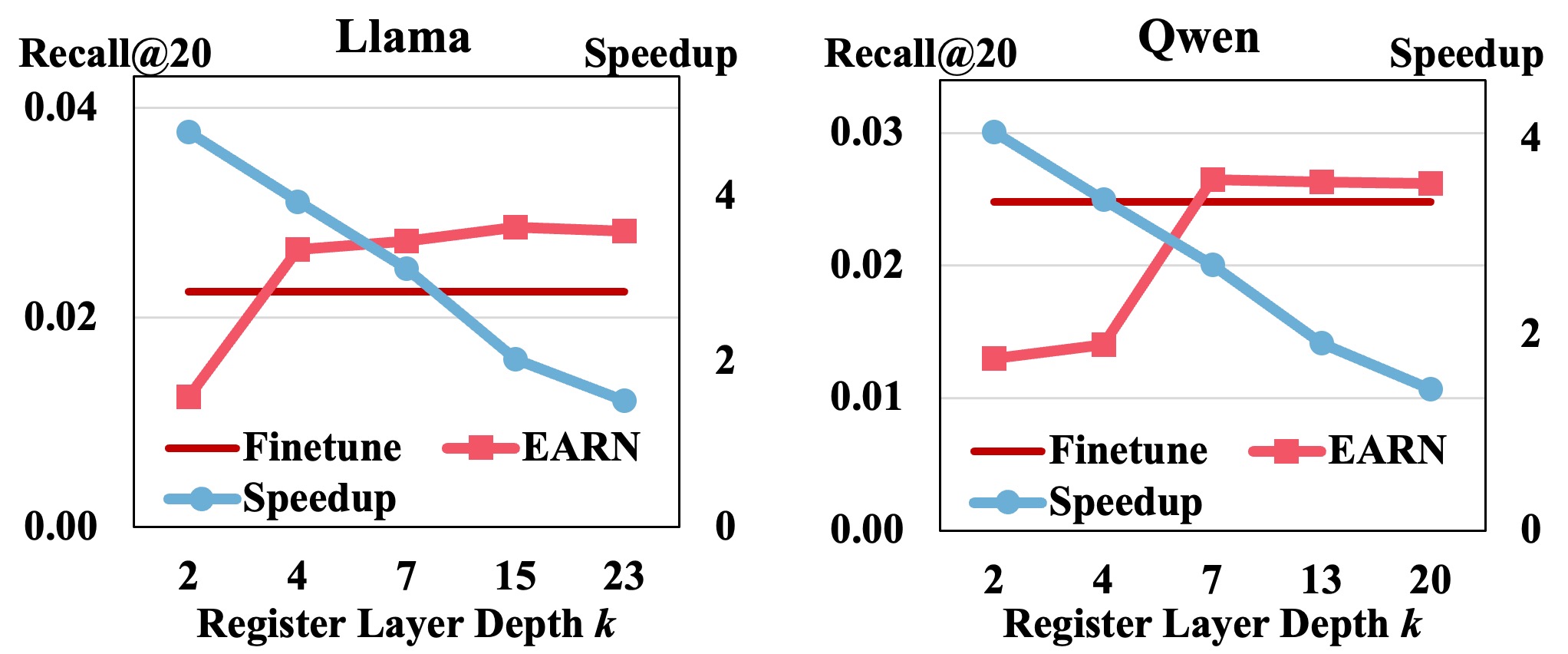}
    \caption{Effect of register layer depth $\bm k$.}
    \label{fig:register_layer}
\end{figure}
Figure~\ref{fig:register_layer} reveals the trade-off between inference efficiency and recommendation effectiveness when varying the register layer depth $k$:
\textbf{1)} Too shallow: When the register layer depth is too shallow (\eg $<4$ on Llama, and $<7$ on Qwen), although the speedup is high, the recall suffers a significant loss.
\textbf{2)} Optimal range: In the range of $k=4-7$, EARN strikes a balance between speedup and recall. The speedup remains reasonably high, while Recall@20 improves notably. This suggests that the model can effectively capture the necessary information for recommendations without a substantial loss in inference efficiency.
\textbf{3)} Too deep: When the register layer depth is too deep (\eg $k \geq 13$), the speedup decreases significantly, but Recall@20 does not show a proportional improvement. For example, on Qwen, the speedup drops sharply when $k=13$ or $k=20$, while Recall@20 does not increase substantially. This indicates that increasing the register layer depth beyond a certain point leads to diminishing returns in terms of recommendation effectiveness while significantly compromising inference efficiency.

\sss{Effect of Register Token Number $\bm n$}  
\begin{figure}[t]
\vspace{-0.2cm}
    \centering
    \setlength{\abovecaptionskip}{0cm}
    \setlength{\belowcaptionskip}{0cm}
    \includegraphics[width=1\linewidth]{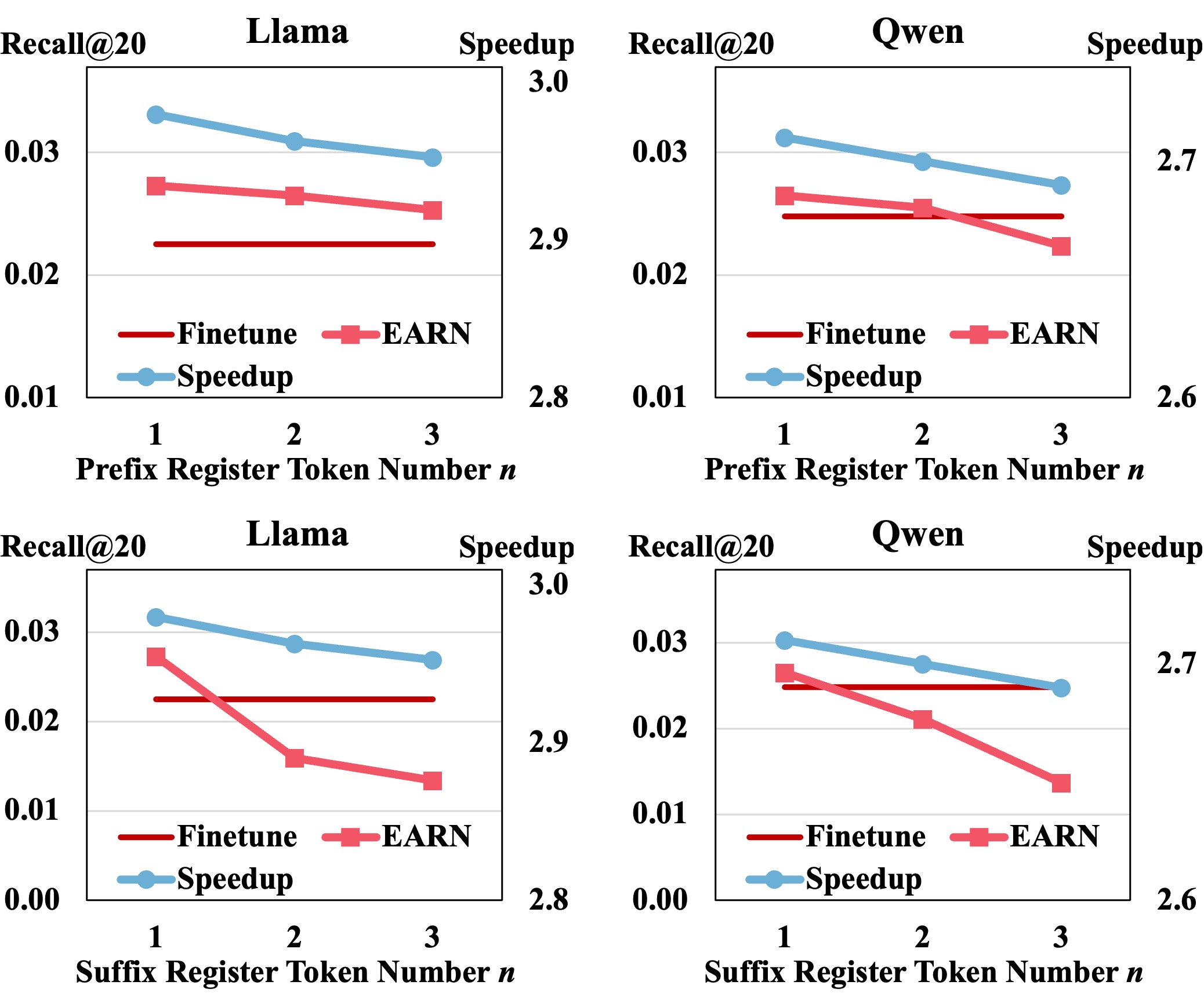}
    \caption{Effect of register token number $\bm n$.}
    \label{fig:register_num}
\end{figure}
Figure~\ref{fig:register_num} reveals the trade-off between inference efficiency and recommendation effectiveness when varying the register token number $n$, from which we can observe that:
\textbf{1)} For both the prefix register and the suffix register, as the register token number $n$ varies from 1 to 3, the speedup gradually decreases. This is primarily because, as the register token number increases, the size of the KV Cache that needs to be computed and cached also grows, leading to increased inference latency.
\textbf{2)} There is a decline in Recall@20 as the register token number $n$ increases for both the prefix register and the suffix register. This phenomenon may be attributed to the interactions among multiple tokens in the later layers of the model, which can lead to a progressive distortion of the summarized information. The impact of this accuracy degradation is relatively minor for the prefix register but more pronounced for the suffix register. This discrepancy is likely due to the fact that user interaction history summarized by the suffix register is more complex and challenging to condense compared to task instructions summarized by the prefix register.

\begin{figure}[t]
\vspace{-0.2cm}
    \centering
    \setlength{\abovecaptionskip}{0cm}
    \setlength{\belowcaptionskip}{0cm}
    \includegraphics[width=1\linewidth]{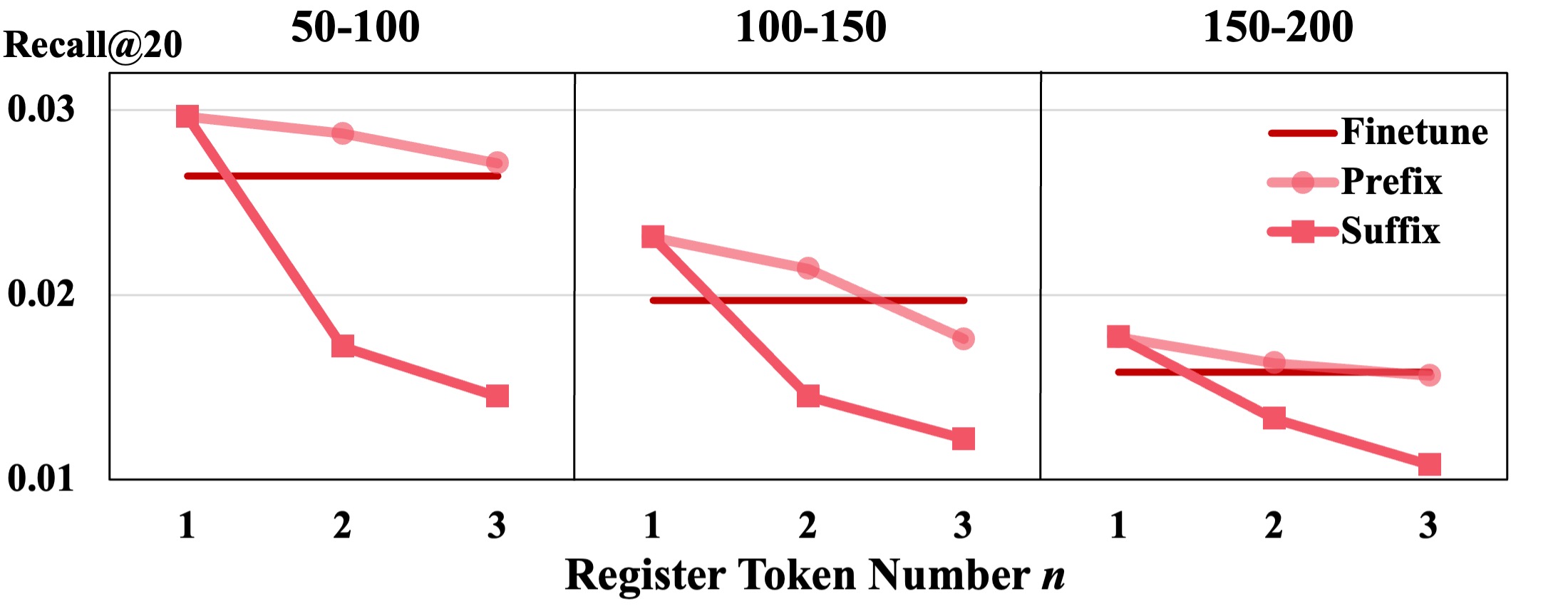}
    \caption{Effect of register token number $\bm n$ under different prompt length.}
    \label{fig:register_num_len}
\end{figure}

Although the above results show that a single register token is optimal, as prompt length increases, the optimal number of register tokens may need to be adjusted to maintain the performance. Therefore, we conduct further experiments to investigate how EARN performs under different prompt lengths and whether the optimal number of register tokens should be adjusted as the prompt length increases. 
We segmented the dataset into three groups based on prompt length (50-100, 100-150, and 150-200 tokens) to evaluate our EARN's performance across different prompt lengths. As shown in Figure~\ref{fig:register_num_len}, EARN consistently outperforms the Finetune baseline, and a single register token remains optimal across all prompt lengths.
However, we observed an interesting trend for the suffix register token: the performance gap between 1 and 2 register tokens narrows as prompt length increases. This suggests that the optimal number of register tokens may rise for very long prompts. 
Nevertheless, experimental results on three real-world datasets (Table~\ref{tab:overall}) demonstrate that employing a single suffix register token could achieve excellent recommendation performance, which is applicable to the majority of recommendation scenarios.

\sss{Hyper-parameter Recommendation}
We experimentally validated that EARN's hyper-parameters can be selected through simple heuristics rather than exhaustive tuning. And we validated that setting the register layer and using a single register token at one-fourth achieves significant acceleration without accuracy loss on three real-world datasets. This configuration is broadly applicable for most recommendation tasks. Detailed analysis is as follows: \textbf{1)} Register layer depth $k$: Firstly, the lower the register layer $k$, the greater the acceleration effect achieved. Secondly, our sensitivity analysis indicates that there is essentially no loss in recommendation effectiveness when $k$ is set to at least one-fourth of the total layers. 
\textbf{2)} Register token number $n$: Across two distinct LLM models (Llama and Qwen), we consistently found that one register token is optimal. Additionally, to assess if the number needs adjustment as prompt length increases, we conducted grouped experiments by length, and found that a single register token remains optimal under different prompt lengths.
Thus, for most LLMRec scenarios, we propose a default configuration:
\begin{enumerate}
    \item \textbf{$\bm k$:} One-fourth of the total layers as the register layer depth.
    \item \textbf{$\bm n$:} One prefix register token and one suffix register token.
\end{enumerate}

This setting could achieve a favorable speedup and KV Cache reduction while maintaining strong recommendation performance. And it’s easily adjustable for various deployment scenarios.

\subsection{Ablation Studies (RQ4)}
To study the contribution of each component of the proposed EARN, we conduct ablation studies on the register training strategy and the register token. 

\sss{Effect of Register Training} 
We first investigate the necessity of register training by comparing it with direct register inference on fully finetuned models. As shown in Table~\ref{tab:ablation_train}, models without RT suffer significant performance degradation across all metrics. Specifically for Llama with $k=7$, removing RT leads to 72\% relative drop in R@10 (from 0.0174 to 0.0048) and 80\% reduction in N@20 (from 0.0127 to 0.0025). This demonstrates that directly applying register inference without dedicated training fails to effectively capture task-specific knowledge.
With RT, both models achieve consistent improvements over Finetune. Llama with $k=7$ gains 17\% higher R@10 (0.0174 vs. 0.0145) and 15\% better N@20 (0.0127 vs. 0.0108). This indicates that register training enables the model to summarize key information from user historical interactions into the register token.

\begin{table}[!t]
    \centering
    \setlength{\abovecaptionskip}{0cm}
    \setlength{\belowcaptionskip}{0cm}
    \caption{Ablation study of the \textit{register training} (RT).}
    \label{tab:ablation_train}
    \setlength{\tabcolsep}{3.5mm}{
\resizebox{0.46\textwidth}{!}{
    \begin{tabular}{ccccccc}
    \toprule
    \textbf{Model} & $\bm k$ & \textbf{Method} & \textbf{R@10} & \textbf{R@20} & \textbf{N@10} & \textbf{N@20} \\ 
    \midrule
    \multirow{5}{*}{\textbf{Llama}} & ~ & Finetune & 0.0145 & 0.0225 & 0.0084 & 0.0108 \\ 
    \cmidrule{2-7}
    ~ & \multirow{2}{*}{\textbf{7}} & EARN & 0.0174 & 0.0273 & 0.0098 & 0.0127 \\ 
    ~ & ~ & w/o RT & 0.0048 & 0.0060 & 0.0021 & 0.0025 \\ 
    \cmidrule{2-7}
    ~ & \multirow{2}{*}{\textbf{15}} & EARN & 0.0168 & 0.0286 & 0.0093 & 0.0127 \\ 
    ~ & ~ & w/o RT & 0.0053 & 0.0083 & 0.0029 & 0.0038 \\ 
    \midrule
    \multirow{5}{*}{\textbf{Qwen}} & ~ & Finetune & 0.0145 & 0.0248 & 0.0087 & 0.0117 \\ 
    \cmidrule{2-7}
    ~ & \multirow{2}{*}{\textbf{7}} & EARN & 0.0155 & 0.0265 & 0.0091 & 0.0122 \\ 
    ~ & ~ & w/o RT & 0.0074 & 0.0093 & 0.0044 & 0.0050 \\ 
    \cmidrule{2-7}
    ~ & \multirow{2}{*}{\textbf{13}} & EARN & 0.0156 & 0.0263 & 0.0098 & 0.0129 \\ 
    ~ & ~ & w/o RT & 0.0072 & 0.0095 & 0.0043 & 0.0050 \\ 
    \bottomrule
    \end{tabular}
    }}
\end{table}

\sss{Effect of Prefix Register and Suffix Register}
We further dissect the contribution of the prefix register and the suffix register through component-wise ablation. Table~\ref{tab:ablation_PR} reveals three key observations: 
\textbf{1)} Suffix register dominates performance: Removing the suffix register (w/o SR) causes significant performance collapse for both Llama and Qwen. This confirms our design intuition that summarizing historical interactions in the suffix register is crucial for recommendation tasks.
\textbf{2)} Prefix register enhances task awareness: Though not as severe as removing the suffix register (w/o SR), removing the prefix register (w/o PR) still leads to a certain degree of performance degradation. This suggests the prefix register effectively primes the model for recommendation task identification.
\textbf{3)} Layer-depth interaction: The impact of removing the suffix register is more pronounced in the lower layers (\eg $k=7$ exhibits a greater performance drop than $k=15$ on Llama), indicating that the suffix register plays a more significant summarization role in the lower layers. The lower layers rely more heavily on the compressed historical interaction information encoded in the suffix register.

\begin{table}[!t]
    \centering
    \setlength{\abovecaptionskip}{0cm}
    \setlength{\belowcaptionskip}{0cm}
    \caption{Ablation study of the \textit{prefix register} (PR) and the \textit{suffix register} (SR).}
    \label{tab:ablation_PR}
    \setlength{\tabcolsep}{3.5mm}{
\resizebox{0.46\textwidth}{!}{
    \begin{tabular}{ccccccc}
    \toprule
    \textbf{Model} & $\bm k$ & \textbf{Method} & \textbf{R@10} & \textbf{R@20} & \textbf{N@10} & \textbf{N@20} \\ 
    \midrule
    \multirow{6}{*}{\textbf{Llama}} & \multirow{3}{*}{\textbf{7}} & EARN & 0.0174 & 0.0273 & 0.0098 & 0.0127 \\ 
    ~ & ~ & w/o PR & 0.0172 & 0.0252 & 0.0108 & 0.0132 \\ 
    ~ & ~ & w/o SR & 0.0041 & 0.0075 & 0.0021 & 0.0031 \\ 
    \cmidrule{2-7}
    ~ & \multirow{3}{*}{\textbf{15}} & EARN & 0.0168 & 0.0286 & 0.0093 & 0.0127 \\ 
    ~ & ~ & w/o PR & 0.0166 & 0.0260 & 0.0103 & 0.0131 \\ 
    ~ & ~ & w/o SR & 0.0074 & 0.0119 & 0.0035 & 0.0047 \\ 
    \midrule
    \multirow{6}{*}{\textbf{Qwen}} & \multirow{3}{*}{\textbf{7}} & EARN & 0.0155 & 0.0265 & 0.0091 & 0.0122 \\ 
    ~ & ~ & w/o PR & 0.0153 & 0.0217 & 0.0097 & 0.0116 \\ 
    ~ & ~ & w/o SR & 0.0044 & 0.0071 & 0.0023 & 0.0031 \\ 
    \cmidrule{2-7}
    ~ & \multirow{3}{*}{\textbf{13}} & EARN & 0.0156 & 0.0263 & 0.0098 & 0.0129 \\ 
    ~ & ~ & w/o PR & 0.0153 & 0.0230 & 0.0102 & 0.0124 \\ 
    ~ & ~ & w/o SR & 0.0065 & 0.0106 & 0.0036 & 0.0049 \\ 
    \bottomrule
    \end{tabular}
    }}
\end{table}

\section{Related Work}

\noindent$\bullet\quad$\textbf{Inference Acceleration of LLMRec}. 
While LLM-based generative recommendations have shown remarkable performance \cite{lin2025can,wang2023generative,wu2024survey,kim2024large,lin2024bridging,lin2024data,li2024survey}, their practical application is hindered by high inference latency \cite{zhou2024survey,li2023large,wu2024efficient_survey}.
To tackle this issue, various research has been proposed. Several techniques leverage knowledge distillation to transfer comprehensible knowledge~\cite{cui2024distillation} or abstract knowledge~\cite{sun2024distillation} from a teacher LLM to a smaller student language model. 
Additionally, some methods apply speculative decoding to achieve lossless decoding acceleration~\cite{lin2024efficient,xi2024decoding}.
Besides, some approaches attempt to design efficient attention mechanisms to reduce computational complexity, such as sparse attention~\cite{fan2021lighter}, linear attention~\cite{liu2023linrec}, and slimming architecture~\cite{lin2025large}.

\noindent$\bullet\quad$\textbf{KV Cache Reduction}. 
It's common practice to utilize KV Cache to reduce redundant computations during LLM inference. However, the use of KV Cache introduces new challenges. KV Cache will increase linearly with the length of the sequence, and the memory required will become larger and larger.
To address this challenge, diverse methods have been proposed to reduce KV Cache, including two primary groups of work, \ie prompt compression and cache compression.
Prompt compression reduces the initial KV Cache size during \textit{prefilling} by shortening the input token sequence. Hard compression methods like SelectiveContext~\cite{li2023compressing} and LLMLingua~\cite{jiang2023llmlingua} filter redundant tokens while preserving natural language syntax, albeit at the cost of fluency. Soft compression techniques, such as AutoCompressor~\cite{chevalier2023adapting} and 500xCompressor~\cite{li2024500xcompressor}, encode prompts into dense latent tokens, achieving higher compression ratios but sacrificing human interpretability.
Cache compression reduces KV cache during \textit{decoding}. 
Existing work employs eviction or merging strategies. Evict-based methods like StreamingLLM~\cite{xiao2023efficient} and SnapKV~\cite{li2024snapkv} selectively evict less important KV Cache through specific rules. 
Merge-based approaches (\eg CAM~\cite{zhangcam} and DMC~\cite{nawrot2024dynamic}) adaptively merge to-be-evicted caches into the remaining ones.

\section{Conclusion}

In this work, we address the critical challenge of inference efficiency in LLM-based recommendation systems (LLMRec), where the massive computational overhead and memory pressure of KV Cache severely hinders practical deployment. 
Through systematic analysis of LLMRec's attention patterns, we identify two pivotal characteristics: 1) layer-wise attention sparsity inversion, where in early layers retain dense informative patterns while later layers exhibit high redundancy, and 2) the dual attention sinks phenomenon, where attention scores concentrate on both head and tail tokens of input sequences.
These insights motivate our proposed EARN method, which introduces prefix and suffix register tokens to compress task instructions and user interaction histories, implementing layer-wise computation pruning.
EARN achieves an 80\% reduction of KV Cache while maintaining essential information integrity. Extensive experiments conducted on three benchmark datasets and two distinct LLM architectures reveal that our EARN attains 3.79x inference acceleration with superior accuracy compared to conventional finetuning approaches. This breakthrough effectively reconciles the longstanding trade-off between inference efficiency and recommendation quality in LLMRec, presenting tangible deployment benefits for industrial-scale recommendation services.
\newpage
\bibliographystyle{ACM-Reference-Format}
\bibliography{main}

\appendix
\section{Appendix}

\subsection{Detailed Analysis of Attention Score Distributions}
\label{app:attention}
In this section, we provide a systematic analysis of attention score distributions in LLMRec through quantitative measurements across two critical dimensions: layer order and token position. To ensure comprehensive insights, we conduct experiments across 13 NLP tasks and three real-world recommendation datasets, two mainstream LLMRec methods, and two distinct LLM architectures. 
Table~\ref{tab:attn_measure} presents the overall results. More detailed quantitative results and visual figures can be found in our GitHub repository\footnote{https://github.com/transcend-0/EARN/blob/main/attentions.md}.
Our findings reveal fundamental differences in attention patterns between NLP tasks and LLMRec tasks, offering critical guidance for designing efficient compression strategies. 

\textbf{Quantitative Measurements of Attention Score Distributions.}
Given a sequence's attention scores $[p_1,p_2,\cdots,p_n]$, 
we test its sparsity by threshold-based sparsity ratio adapted from prior studies~\cite{deng2025sparseattentionapproximatesexact, chen2021scatterbrain}:
\begin{equation}
    Sparsity = \frac{1}{n} \sum_{i=1}^n \mathbb{I}(p_i > \epsilon)\ ,
\end{equation}
where $\epsilon$ is the threshold. Here we set $\epsilon=0.05$, according to the distribution characteristics of attention scores.

We test its sink by position-based total attention scores:
\begin{equation}
    Sink_{head} = \sum_{i=1}^{T_{h}} p_i \ ,\quad
    Sink_{tail} = \sum_{i=T_{t}}^{n} p_i \ ,
\end{equation}
where $T_{h}$ and $T_{t}$ is the position of the head and tail respectively. Here we set $T_{h}=3, T_{t}=n-3$, according to the distribution characteristics of attention scores.

\textbf{Attention Differences across Layer Orders.}
As measured in Table~\ref{tab:attn_measure}, we identify a \textbf{\textit{layer-wise attention sparsity inversion}} between NLP tasks and LLMRec tasks. 
In NLP tasks, the attention sparsity is high in early layers, but decreases in later layers ($Sp_{early} > Sp_{latter}$). 
Conversely, LLMRec tasks display an inverted pattern: the attention sparsity is relatively low in early layers, but increases in later layers ($Sp_{early} < Sp_{latter}$). 
This indicates that the less sparse early layers in LLMRec retain more user preference information, while high sparsity in later layers suggests redundancy.

\textbf{Attention Differences across Token Positions.}  
Table~\ref{tab:attn_measure} reveals a \textbf{\textit{dual attention sinks phenomenon}} that distinguishes LLMRec from NLP tasks. 
There is significant $Sink_{head}$ for both tasks, while LLMRec exhibits a larger $Sink_{tail}$ ($Sink_{tail}(\text{LLMRec}) > Sink_{tail}(\text{NLP})$). This indicates that both head and tail tokens in LLMRec are crucial for recommendation performance.

\textbf{Implications for Compression Strategy.}
These findings indicate that in LLMRec, the early layers contain richer information, while the middle tokens in the later layers are redundant. 
Our approach takes into account the unique attention score distributions in LLMRec and formulates a reasonable compression scheme from a global perspective. It focuses on preserving information from critical token positions and early layers while pruning redundant parts in later layers. This ensures that it can improve efficiency while maintaining accuracy.

\begin{table}[t]
    \centering
    \setlength{\abovecaptionskip}{0cm}
    \setlength{\belowcaptionskip}{0cm}
    \caption{Measurements of attention sink and sparsity (averaged across 13 NLP tasks, 3 real-world recommendation datasets, 2 LLMRec backbones, and 2 LLM models). Here, $\bm{Sp_{early}}$ denotes  $\bm{Sparsity_{early\ layers}}$, and $\bm{Sp_{latter}}$ denotes $\bm{Sparsity_{latter\ layers}}$.}
    \fontsize{7pt}{7.55pt}\selectfont
    \label{tab:attn_measure}
    \begin{tabular}{cccccc}
    \toprule
    \textbf{Model} & \textbf{Task} & $\bm{Sp_{early}}$ & $\bm{Sp_{latter}}$ & $\bm{Sink_{head}}$ & $\bm{Sink_{tail}}$ \\ \midrule
    \multirow{2}{*}{\textbf{Llama}} & NLP & 0.064 & 0.026 & 0.73 & 0.07 \\
    ~ & LLMRec & 0.025 & 0.048 & 0.56 & 0.09 \\ \midrule
    \multirow{2}{*}{\textbf{Qwen}} & NLP & 0.074 & 0.046 & 0.30 & 0.15 \\
    ~ & LLMRec & 0.047 & 0.059 & 0.30 & 0.24 \\
    \bottomrule
    \end{tabular}
\end{table}

\subsection{Computational Complexity of LLM}
\label{app:inference_speed}

For a decoder-only LLM with $N$ decoder layers, the computational operations comprise three components: embedding, $N$ stacked decoder layers, and de-embedding. Since embedding and de-embedding mainly involve lightweight lookup and projection steps, the computational bottleneck lies in the decoder layers. 
Specifically, each decoder layer consists of two core components: Multi-Head Attention (MHA) and Feed-Forward Network (FFN), whose FLOPs are analyzed as follows:

\textbf{Multi-Head Attention (MHA).} 
Let $L$ denote the sequence length, $d_h$ the hidden dimension, $d_a$ the attention head dimension, and $n_h$ the number of attention heads. We approximate the FLOPs of matrix multiplication $\bm{X}^{L \times d_h} \times \bm{W}^{d_h \times d_a}$ as $2Ld_hd_a$.
For a single attention head, the FLOPs include:  
\textbf{1)} Projections for $\bm{Q}$, $\bm{K}$ and $\bm{V}$: $2Ld_hd_a \times 3 = 6Ld_hd_a$; 
\textbf{2)} Attention computation for $\bm{QK}^T$ and $(\cdot)\bm{V}$: $2L^2d_a + 2L^2d_a = 4L^2d_a$.  
Aggregating across $n_h$ heads, plus $2Ln_hd_ad_h$ FLOPs of the final output projection, the total MHA FLOPs per decoder layer are:  
\begin{equation}
    FLOPs_{MHA} = n_h(8Ld_hd_a + 4L^2d_a)\ .
\end{equation}

\textbf{Feed-Forward Network (FFN).} 
The FFN module involves two projections with intermediate dimension $d_f$:  
\textbf{1)} Up-projection: $2Ld_hd_f$;   
\textbf{2)} Down-projection: $2Ld_fd_h$.  
This yields total FFN FLOPs of:  
\begin{equation}
    FLOPs_{FFN} = 4Ld_hd_f\ .
\end{equation}

\textbf{Total FLOPs.}  
Combining both components, the FLOPs for a single decoder layer are:
\begin{equation}
    \begin{aligned}
    FLOPs_{LLM\ layer} &= n_h(8Ld_hd_a + 4L^2d_a) + 4Ld_hd_f \\
    &= 4L[n_h d_a(2d_h + L)+d_h d_f]\ .
    \end{aligned}
\end{equation}

\subsection{Experimential Details}
\textbf{Datasets Details.} 
\label{app:dataset}
For all three datasets, all historical interactions are sorted according to the global timestamps, and then split into training, validation, and testing sets with the ratio of 8:1:1. 
For the item identifier, we follow LC-Rec~\cite{zheng2024adapting} and TIGER~\cite{rajput2023recommender} to set the length $L=4$, \ie the token sequence length of a generated item would be 4.

\textbf{Implementation details.}
All training and inference experiments are conducted on a single NVIDIA H100 80GB GPU. For the training setup, we employ full finetuning with AdamW optimizer and an overall batch size of $128$ by leveraging gradient accumulation. The learning rate is set to $0.001$, and is adjusted dynamically by a cosine learning rate scheduler with a warmup ratio of $0.02$.
For the inference setup, we set the beam size to $20$ and use the maximum batch size that the GPU could accommodate. 
For the hyper-parameters of EARN in Table 1, EARN on Llama uses $k = 4$ and $n=1$, and EARN on Qwen uses $k = 7$ and $n=1$.

\subsection{Additional Results on TIGER}
\label{app:TIGER}
Table~\ref{tab:TIGER} shows the overall performance comparison between the baselines and EARN instantiated on TIGER. Our EARN achieves the best performance in terms of time efficiency and recommendation effectiveness, while also demonstrating excellent space efficiency. These results further validate the effectiveness of EARN.

\subsection{Additional Results on HSTU}
We also conduct experiments on HSTU~\cite{zhai2024actions}, the first industry-deployed generative recommendation system. As shown in Table~\ref{tab:HSTU}, our EARN is also effective for HSTU, demonstrating excellent time and space efficiency (2x speedup and 50\% memory savings), with minimal impact on recommendation effectiveness.
It is important to note that HSTU differs from other popular LLM-based generative recommendation methods (\eg LC-Rec). 
Specifically, in HSTU model, there are no textual inputs and no concept of prompts as seen in NLP tasks. And HSTU directly generates the predicted item embeddings, akin to generating only one token, which means it only has the prefilling stage and no decoding stage.
These render that the baselines of prompt compression, cache compression, and gist methods in NLP are not applicable.
Therefore, we primarily compared Finetune, SkipLayer, and EARN.

\subsection{Additional Results on NLP Tasks}

To validate the generalizability of EARN beyond recommendation tasks, we conducted experiments on MMLU dataset~\cite{hendryckstest2021}—a general NLP dataset spanning 57 tasks (e.g., math, science, and humanities).

Results in Table~\ref{tab:MMLU} demonstrate EARN's effectiveness:
\textbf{1)} Efficiency-accuracy trade-off: As the register layer $k$ decreases, EARN achieves higher speedup and cache reduction at the cost of reduced accuracy. However, when $k\ge15$, the accuracy loss remains within 10\%, indicating that EARN can still deliver satisfactory performance for general NLP tasks. 
\textbf{2)} Comparison with LLMRec tasks: Unlike LLMRec tasks, where EARN incurs no accuracy loss at $k\ge 4$ and even boosts accuracy by pruning noisy information (refer to Figure~\ref{fig:register_layer}), NLP tasks require a higher $k$ to curb accuracy degradation. 

In summary, while optimal $k$ varies across tasks, it consistently enables efficient inference with controlled accuracy trade-offs, showcasing its broader applicability beyond recommendation tasks.

\begin{table}[!t]
    \centering
    \setlength{\abovecaptionskip}{0cm}
    \setlength{\belowcaptionskip}{0cm}
    \caption{Overall performance comparison between the baselines and EARN instantiated on TIGER.
    }
    \label{tab:TIGER}
    \setlength{\tabcolsep}{0.7mm}
    \fontsize{7pt}{7.55pt}\selectfont
    \begin{tabular}{@{}l|rr|rr|ccccc@{}}
    \toprule
    \multirow{2}{*}{\textbf{Method}} & \multicolumn{2}{c|}{\fontsize{6pt}{6pt}\selectfont\textbf{Time Efficiency}} & \multicolumn{2}{c|}{\fontsize{6pt}{6pt}\selectfont\textbf{Space Efficiency}} & \multicolumn{4}{c}{\fontsize{6pt}{6pt}\selectfont\textbf{Recommendation Effectiveness}} \\
     & $\bm\omega$ & $\bm\tau$ & $\bm\gamma$ & $\bm\sigma$ & \textbf{R@10} & \textbf{R@20} & \textbf{N@10} & \textbf{N@20} \\
    \midrule \multicolumn{5}{@{}l}{\textit{\textbf{Beauty, Llama}}} \\ \midrule
    \textbf{Finetune} & 1.00 & 602.50 & 0.0 & 68.18 & \uline{0.0108} & \uline{0.0198} & \uline{0.0071} & \uline{0.0096} \\ 
    \textbf{SkipLayers} & 1.78 & 1069.7 & 44.5 & 37.87 & 0.0010 & 0.0010 & 0.0011 & 0.0011 \\ 
    \textbf{POD} & 1.17 & 701.7 & 14.8 & 58.11 & 0.0033 & 0.0065 & 0.0027 & 0.0039 \\ 
    \textbf{500xCompressor} & \uline{2.31} & \uline{1389.1} & 74.8 & 17.16 & 0.0003 & 0.0003 & 0.0001 & 0.0001 \\ 
    \textbf{StreamingLLM} & 1.21 & 730.0 & \textbf{96.4} & \textbf{2.44} & 0.0003 & 0.0005 & 0.0005 & 0.0003 \\ 
    \textbf{SnapKV} & 1.19 & 718.1 & \uline{94.5} & \uline{3.76} & 0.0041 & 0.0053 & 0.0027 & 0.0031 \\ 
    \cellcolor{gray!16}\textbf{EARN} & \cellcolor{gray!16}\textbf{3.44} & \cellcolor{gray!16}\textbf{1972.4} & \cellcolor{gray!16}80.9 & \cellcolor{gray!16}13.04 & \cellcolor{gray!16}\textbf{0.0115} & \cellcolor{gray!16}\textbf{0.0199} & \cellcolor{gray!16}\textbf{0.0075} & \cellcolor{gray!16}\textbf{0.0098} \\ 
    \midrule \multicolumn{5}{@{}l}{\textit{\textbf{Beauty, Qwen}}} \\ \midrule
    \textbf{Finetune} & 1.00 & 714.3 & 0.0 & 12.72 & \uline{0.0095} & \uline{0.0162} & \uline{0.0061} & \uline{0.0081} \\ 
    \textbf{SkipLayers} & 1.69 & 1208.1 & 57.9 & 5.35 & 0.0000 & 0.0000 & 0.0000 & 0.0000 \\ 
    \textbf{POD} & 1.10 & 784.4 & 8.4 & 11.65 & 0.0056 & 0.0084 & 0.0032 & 0.0044 \\ 
    \textbf{500xCompressor} & \uline{2.55} & \uline{1819.2} & \uline{91.4} & \uline{1.09} & 0.0003 & 0.0005 & 0.0002 & 0.0005 \\ 
    \textbf{StreamingLLM} & 1.05 & 751.6 & \textbf{91.7} & \textbf{1.06} & 0.0057 & 0.0096 & 0.0040 & 0.0052 \\ 
    \textbf{SnapKV} & 1.02 & 728.7 & 69.6 & 3.87 & 0.0063 & 0.0110 & 0.0043 & 0.0055 \\ 
    \cellcolor{gray!16}\textbf{EARN} & \cellcolor{gray!16}\textbf{2.58} & \cellcolor{gray!16}\textbf{1805.9} & \cellcolor{gray!16}75.5 & \cellcolor{gray!16}3.11 & \cellcolor{gray!16}\textbf{0.0104} & \cellcolor{gray!16}\textbf{0.0164} & \cellcolor{gray!16}\textbf{0.0065} & \cellcolor{gray!16}\textbf{0.0089} \\ 
    \midrule \multicolumn{5}{@{}l}{\textit{\textbf{Games, Llama}}} \\ \midrule
    \textbf{Finetune} & 1.00 & 695.8 & 0.0 & 61.40 & \uline{0.0126} & \textbf{0.0216} & \uline{0.0080} & \uline{0.0107} \\ 
    \textbf{SkipLayers} & 1.83 & 1272.4 & 45.0 & 33.76 & 0.0001 & 0.0001 & 0.0003 & 0.0003 \\ 
    \textbf{POD} & 1.08 & 757.8 & 15.8 & 51.67 & 0.0054 & 0.0098 & 0.0032 & 0.0045 \\ 
    \textbf{500xCompressor} & \uline{2.11} & \uline{1470.2} & 72.5 & 16.88 & 0.0010 & 0.0015 & 0.0007 & 0.0008 \\ 
    \textbf{StreamingLLM} & 1.22 & 850.7 & \textbf{96.0} & \textbf{2.46} & 0.0010 & 0.0010 & 0.0010 & 0.0009 \\ 
    \textbf{SnapKV} & 1.21 & 837.6 & \uline{93.5} & \uline{3.98} & 0.0064 & 0.0075 & 0.0045 & 0.0045 \\ 
    \cellcolor{gray!16}\textbf{EARN} & \cellcolor{gray!16}\textbf{3.12} & \cellcolor{gray!16}\textbf{2062.7} & \cellcolor{gray!16}81.2 & \cellcolor{gray!16}11.56 & \cellcolor{gray!16}\textbf{0.0138} & \cellcolor{gray!16}\uline{0.0212} & \cellcolor{gray!16}\textbf{0.0085} & \cellcolor{gray!16}\textbf{0.0108} \\ 
    \midrule \multicolumn{5}{@{}l}{\textit{\textbf{Games, Qwen}}} \\ \midrule
    \textbf{Finetune} & 1.00 & 673.2 & 0.0 & 10.81 & \uline{0.0131} & \uline{0.0224} & \uline{0.0085} & \uline{0.0113} \\ 
    \textbf{SkipLayers} & 1.51 & 1014.3 & 57.0 & 4.65 & 0.0000 & 0.0000 & 0.0000 & 0.0000 \\ 
    \textbf{POD} & 1.36 & 912.1 & 8.5 & 9.88 & 0.0087 & 0.0150 & 0.0050 & 0.0073 \\ 
    \textbf{500xCompressor} & \uline{2.60} & \uline{1752.0} & \uline{96.7} & \uline{0.35} & 0.0002 & 0.0004 & 0.0001 & 0.0001 \\ 
    \textbf{StreamingLLM} & 1.06 & 708.8 & \textbf{98.4} & \textbf{0.17} & 0.0088 & 0.0145 & 0.0050 & 0.0072 \\ 
    \textbf{SnapKV} & 1.02 & 686.4 & 67.6 & 3.50 & 0.0094 & 0.0161 & 0.0056 & 0.0081 \\ 
    \cellcolor{gray!16}\textbf{EARN} & \cellcolor{gray!16}\textbf{2.86} & \cellcolor{gray!16}\textbf{1841.3} & \cellcolor{gray!16}72.1 & \cellcolor{gray!16}3.02 & \cellcolor{gray!16}\textbf{0.0137} & \cellcolor{gray!16}\textbf{0.0231} & \cellcolor{gray!16}\textbf{0.0086} & \cellcolor{gray!16}\textbf{0.0119} \\ 
    \midrule \multicolumn{5}{@{}l}{\textit{\textbf{MovieLens, Llama}}} \\ \midrule
    \textbf{Finetune} & 1.00 & 852.50 & 0.0 & 39.79 & \uline{0.0245} & \uline{0.0446} & \uline{0.0244} & \uline{0.0323} \\ 
    \textbf{SkipLayers} & 1.85 & 1577.0 & 67.6 & 12.89 & 0.0022 & 0.0024 & 0.0059 & 0.0051 \\ 
    \textbf{POD} & 1.18 & 1003.9 & 18.1 & 32.57 & 0.0068 & 0.0122 & 0.0088 & 0.0103 \\ 
    \textbf{500xCompressor} & \uline{1.93} & \uline{1640.0} & 61.1 & 15.48 & 0.0004 & 0.0006 & 0.0004 & 0.0003 \\ 
    \textbf{StreamingLLM} & 1.18 & 1007.9 & \textbf{95.2} & \textbf{1.89} & 0.0000 & 0.0000 & 0.0000 & 0.0000 \\ 
    \textbf{SnapKV} & 1.18 & 1002.50 & \uline{91.8} & \uline{3.28} & 0.0000 & 0.0000 & 0.0000 & 0.0000 \\ 
    \cellcolor{gray!16}\textbf{EARN} & \cellcolor{gray!16}\textbf{2.85} & \cellcolor{gray!16}\textbf{2437.4} & \cellcolor{gray!16}77.7 & \cellcolor{gray!16}8.87 & \cellcolor{gray!16}\textbf{0.0269} & \cellcolor{gray!16}\textbf{0.0544} & \cellcolor{gray!16}\textbf{0.0221} & \cellcolor{gray!16}\textbf{0.0340} \\ 
    \midrule \multicolumn{5}{@{}l}{\textit{\textbf{MovieLens, Qwen}}} \\ \midrule
    \textbf{Finetune} & 1.00 & 897.4 & 0.0 & 7.75 & \uline{0.0238} & \uline{0.0455} & \uline{0.0224} & \uline{0.0312} \\ 
    \textbf{SkipLayers} & 1.49 & 1340.5 & 76.9 & 1.79 & 0.0002 & 0.0002 & 0.0002 & 0.0002 \\ 
    \textbf{POD} & 1.41 & 1260.9 & 39.8 & 4.66 & 0.0114 & 0.0192 & 0.0106 & 0.0131 \\ 
    \textbf{500xCompressor} & \uline{2.10} & \uline{1882.1} & \uline{77.9} & \uline{1.71} & 0.0004 & 0.0009 & 0.0005 & 0.0007 \\ 
    \textbf{StreamingLLM} & 1.01 & 912.2 & \textbf{88.3} & \textbf{0.91} & 0.0213 & 0.0402 & 0.0180 & 0.0241 \\ 
    \textbf{SnapKV} & 0.97 & 872.4 & 76.5 & 1.82 & 0.0204 & 0.0464 & 0.0174 & 0.0263 \\ 
    \cellcolor{gray!16}\textbf{EARN} & \cellcolor{gray!16}\textbf{2.56} & \cellcolor{gray!16}\textbf{2292.8} & \cellcolor{gray!16}45.5 & \cellcolor{gray!16}4.23 & \cellcolor{gray!16}\textbf{0.0274} & \cellcolor{gray!16}\textbf{0.0474} & \cellcolor{gray!16}\textbf{0.0273} & \cellcolor{gray!16}\textbf{0.0355} \\ 
    \bottomrule
    \end{tabular}
\end{table}

\begin{table}[h]
    \centering
    \setlength{\abovecaptionskip}{0cm}
    \setlength{\belowcaptionskip}{0cm}
    \caption{Overall comparison between baselines and EARN instantiated on HSTU. Here, the unit of $\bm\sigma$ is MB.}
    \label{tab:HSTU}
    \setlength{\tabcolsep}{0.7mm}
    \fontsize{7pt}{7.55pt}\selectfont
    \begin{tabular}{@{}l|rr|rr|cccc@{}}
    \toprule
    \multirow{2}{*}{\textbf{Method}} & \multicolumn{2}{c|}{\fontsize{6pt}{6pt}\selectfont\textbf{Time Efficiency}} & \multicolumn{2}{c|}{\fontsize{6pt}{6pt}\selectfont\textbf{Space Efficiency}} & \multicolumn{4}{c}{\fontsize{6pt}{6pt}\selectfont\textbf{Recommendation Effectiveness}} \\
    & $\bm\omega$ & $\bm\tau$ & $\bm\gamma$ & $\bm\sigma$ & \textbf{R@10} & \textbf{R@20} & \textbf{N@10} & \textbf{N@20} \\
    \midrule \multicolumn{5}{@{}l}{\textit{\textbf{Beauty}}} \\ \midrule
    \textbf{Finetune} & 1.00 & 2735.5 & 0.0 & 439.5 & \uline{0.0446} & \textbf{0.0693} & \uline{0.0269} & \uline{0.0328} \\
    \textbf{SkipLayers} & \uline{1.72} & \uline{4685.6} & \uline{27.8} & \uline{317.5} & 0.0248 & 0.0322 & 0.0116 & 0.0136 \\
    \cellcolor{gray!16}\textbf{EARN} & \cellcolor{gray!16}\textbf{2.01} & \cellcolor{gray!16}\textbf{5414.9} & \cellcolor{gray!16}\textbf{49.5} & \cellcolor{gray!16}\textbf{222.0} & \cellcolor{gray!16}\textbf{0.0520} & \cellcolor{gray!16}\uline{0.0594} & \cellcolor{gray!16}\textbf{0.0311} & \cellcolor{gray!16}\textbf{0.0329} \\
    \midrule \multicolumn{5}{@{}l}{\textit{\textbf{Games}}} \\ \midrule
    \textbf{Finetune} & 1.00 & 3425.2 & 0.0 & 532.9 & \uline{0.0557} & \textbf{0.0846} & \uline{0.0362} & \textbf{0.0435} \\
    \textbf{SkipLayers} & \uline{1.69} & \uline{5797.2} & \uline{27.9} & \uline{384.4} & 0.0371 & 0.0520 & 0.0200 & 0.0237 \\
    \cellcolor{gray!16}\textbf{EARN} & \cellcolor{gray!16}\textbf{1.97} & \cellcolor{gray!16}\textbf{6770.9} & \cellcolor{gray!16}\textbf{50.3} & \cellcolor{gray!16}\textbf{264.9} & \cellcolor{gray!16}\textbf{0.0586} & \cellcolor{gray!16}\uline{0.0735} & \cellcolor{gray!16}\textbf{0.0384} & \cellcolor{gray!16}\uline{0.0421} \\
    \midrule \multicolumn{5}{@{}l}{\textit{\textbf{MovieLens}}} \\ \midrule
    \textbf{Finetune} & 1.00 & 3187.9 & 0.0 & 546.9 & 0.0756 & 0.1073 & 0.0405 & 0.0485 \\
    \textbf{SkipLayers} & \uline{1.75} & \uline{5602.4} & \uline{34.6} & \uline{357.7} & \uline{0.0780} & \uline{0.1207} & \uline{0.0429} & \uline{0.0536} \\
    \cellcolor{gray!16}\textbf{EARN} & \cellcolor{gray!16}\textbf{2.06} & \cellcolor{gray!16}\textbf{6555.8} & \cellcolor{gray!16}\textbf{50.5} & \cellcolor{gray!16}\textbf{270.6} & \cellcolor{gray!16}\textbf{0.0793} & \cellcolor{gray!16}\textbf{0.1268} & \cellcolor{gray!16}\textbf{0.0429} & \cellcolor{gray!16}\textbf{0.0551} \\
    \bottomrule
    \end{tabular}
\end{table}

\begin{table}[!t]
    \centering
    \setlength{\abovecaptionskip}{0cm}
    \setlength{\belowcaptionskip}{0cm}
    \caption{Overall performance comparison between Finetune and EARN with different register layer $k$ on Llama (total 32 layers) on MMLU.}
    \label{tab:MMLU}
    \fontsize{7pt}{7.55pt}\selectfont
    \begin{tabular}{c|rr|rr|c}
    \toprule
    \textbf{RegisterLayer $\bm k$} & $\bm\omega$ & $\bm\tau$ & $\bm\gamma$ & $\bm\sigma$ & \textbf{Accuracy} \\ 
    \midrule
    \textbf{Finetune} & 1.0 & 76.0 & 0.0 & 266.0 & \textbf{0.51} \\
    \textbf{4} & \textbf{6.9} & \textbf{523.6} & \textbf{83.3} & \textbf{44.5} & 0.27 \\
    \textbf{7} & \uline{4.3} & \uline{331.3} & \uline{73.3} & \uline{71.0} & 0.28 \\
    \textbf{11} & 2.9 & 219.0 & 61.3 & 103.0 & 0.28 \\
    \textbf{15} & 2.2 & 163.4 & 49.3 & 135.0 & 0.46 \\
    \textbf{19} & 1.7 & 130.3 & 38.2 & 164.5 & 0.50 \\
    \textbf{23} & 1.4 & 104.3 & 26.1 & 196.5 & \uline{0.50} \\
    \textbf{27} & 1.2 & 91.6 & 14.2 & 228.3 & 0.50 \\
    \bottomrule
    \end{tabular}
\end{table}

\end{document}